\documentstyle[12pt,epsfig]{article}
\def\la{\mathrel{\mathpalette\fun <}}
\def\ga{\mathrel{\mathpalette\fun >}}
\def\fun#1#2{\lower3.6pt\vbox{\baselineskip0pt\lineskip.9pt
\ialign{$\mathsurround=0pt#1\hfil##\hfil$\crcr#2\crcr\sim\crcr}}}

\begin{document}

\begin{titlepage}
\vspace*{1cm}
\hspace{10cm} CMS NOTE 1997/094

\hspace{10cm} HEPHY-PUB 678/97  
\begin{center}

\vspace{0.5cm}

{\Large \bf Determining  the parameters of the Minimal Supergravity
Model from \boldmath{$2l + E_T^{miss} + (jets)$} final states at LHC.}

\end{center}

\vspace{1.5cm}

\begin{center} 
         D.~Denegri$^1$, W.Majerotto$^2$, and L.~Rurua$^{2,3}$

\vspace{1cm}
 $^1${\it Centre d'Etudes Nucle$\acute{a}$ire de Saclay, Gif-sur-Yvette, 
France}

\vspace{0.3cm}  
 $^2${\it Institut f$\ddot{u}$r Hochenergiephysik, 
$\ddot{O}$sterreichische Akademie d.Wissenschaften, Vienna, Austria}

\vspace{0.3cm}
 $^3${\it Institute of Physics, Tbilisi, Georgia} 
\end{center}

\vspace{3cm}

\begin{abstract}

We analyse the events with two same--flavour, opposite--sign leptons
+ $E_T^{\rm{miss}}$ + (jets) as expected in $pp$ collisions at LHC
within the framework of the minimal Supergravity Model. The
objective is the determination of the parameters $m_0$ and  $m_{1/2}$
of this model (for a given value of $\tan\beta$). The
signature $l^+ l^- + E_T^{miss}$ + (jets)  selects the leptonic decays of 
$\tilde{\chi}^0_2$, $\tilde{\chi}^0_2 \rightarrow \tilde{\chi}^0_1 l^+ l^- $, 
$\tilde{\chi}^0_2 \rightarrow \tilde{l}_{L,R}^{\pm} l^{\mp}
\rightarrow \tilde{\chi}^0_1 l^+ l^-$. 
We exploit the fact that the invariant dilepton mass distribution has 
a pronounced structure with a sharp edge at the kinematical endpoint 
even in such an inclusive final state over a significant part of parameter 
space. We determine the domain of parameter space where the edge is expected 
to be visible. 
We show that a measurement of this edge already constrains the model
parameters essentially 
to three lines in the ($m_0, m_{1/2}$) parameter plane. We work
out a strategy to discriminate between the 
three--body leptonic decays of $\tilde{\chi}^0_2$ and the 
decays into sleptons
$\tilde{l}_{L,R}$. This procedure may make it 
possible to get information on SUSY particle masses already with low
luminosity, $L_{int}=10^3$ pb$^{-1}$.

\end{abstract}
\end{titlepage}

\section{Introduction} 
If 'low-energy' supersymmetry (SUSY) is realised in Nature it should
show up at the Large Hadron Collider (LHC). Strongly interacting
particles  as gluinos and squarks will be most likely the first SUSY
particles to be seen at LHC. Gluinos of mass less than $\sim 2$ TeV
and  squarks of mass less  than $\sim 1.5$ TeV [1-3] can be detected,
covering in such a way the whole  theoretically motivated parameter
space. LHC is also a good laboratory for the search of electro-weakly
interacting particles, e.g. sleptons [4,5]. In a  recent paper [6] it
was shown within the minimal supergravity (mSUGRA) [7] model that
sleptons in the mass range of  $\sim 100$ to 400 GeV can be  detected
at LHC by investigating the signature $two\,\, leptons + E_T^{miss} +
no \,\,jets$. However, this final state where direct production (Drell-Yan) 
of sleptons predominates requires high luminosity, $L_{int}=10^5\,\, pb^{-1}$. 
Sleptons can be also produced indirectly
in the decays of charginos and neutralinos, especially in
$\tilde{\chi}_2^0 \rightarrow \tilde{l}_{L,R} l$ decays.
The charginos and neutralinos, can in turn be produced directly or come
from gluinos and/or squarks.  This leads to final states with
$ \geq 2 \, leptons + E_T^{miss} + (jets)$.
Actually, this indirect slepton production through $\tilde{g},\tilde{q}$ 
decays has the largest cross-section in a sizable region
of the parameter space accessible at LHC and could allow sleptons to be
already revealed at $L_{int}=10^3 \,\,pb^{-1}$, i.e. simultaneously
with strongly interacting sparticles. Thus, indirect production of
sleptons can be more important for observing a slepton signal
than direct one [8]. Moreover, 
in such a way the mass reach for sleptons search can be extended up to
$m_{\tilde{l}_L}\sim $ 740 GeV. 

Having evidence for SUSY at LHC, one of the next tasks will be to find out
the underlying model and to determine the model parameters.
In this paper we work out a method to determine SUSY parameters and 
suggest a strategy for getting information on masses of SUSY particles 
by means of the signature  $l^+l^- + E_T^{miss} + (jets)$.
Our study is made within the framework of the minimal supergravity
model (mSUGRA) [7]. In this model all scalar
particles (sfermions and Higgs bosons) have a common mass $m_0$ at
$M_{\rm{GUT}} \approx 10^{16}$~GeV. The gaugino masses $M_1 , M_2 ,
M_3$ (corresponding to U(1), SU(2), and SU(3), respectively) unify to a
common gaugino mass $m_{1/2}$, and all trilinear coupling parameters
$A_{ijk}$ have the same value $A_0$ at $M_{\rm{GUT}}$. One also has
unification of the electroweak and strong coupling parameters
$\alpha_i , i = 1,2,3$ [9]. A further reduction of the parameters is given
by invoking 'radiative symmetry breaking'. As a consequence, one has
only the following input parameters: $m_0 , m_{1/2} , A_0 , \tan\beta,
{\rm{sign}} (\mu )$. Here $\tan\beta = \frac{v_2}{v_1}$, the ratio
of the two vacuum expectation values of the two Higgs doublets, and
$\mu$ is the Higgsino mass parameter. The whole SUSY particle spectrum
can then be calculated by making use of renormalization group
equations (RGE).  This model is also incorporated in the Monte--Carlo
generator ISASUSY [10] which is used in our analysis.

This paper is aimed at determining the parameters $m_0$ and $m_{1/2}$ 
with fixed $tan\beta$. Knowing these parameters, we can calculate the masses 
of the superpartners using RGE. 
For this purpose, we study in detail the leptonic decays of  
$\tilde{\chi}_2^0$ which have some useful features. 
Within the mSUGRA model, $\tilde{\chi}_2^0$ has two-body decays,
$\tilde{\chi}_2^0\rightarrow \tilde{l}_{L,R}^{\pm}l^{\mp}$,
in the region $m_0 \la 0.5 \cdot m_{1/2}$ of the parameter space, 
whereas in the region $m_0 \ga 0.5 \cdot m_{1/2}$, $m_{1/2}
\la 200$ GeV the  $\tilde{\chi}_2^0$  has three-body decays,
$\tilde{\chi}_2^0\rightarrow l^+l^-\tilde{\chi}_1^0$. In 
both regions the invariant dilepton mass spectrum ($M_{l^+l^-}$) has 
a maximum $M_{l^+l^-}^{max}$, and therefore a pronounced structure 
with a sharp edge can be seen at the kinematical endpoint.
This property was discussed first in ref. [11] in the case of three-body 
decays of $\tilde{\chi}_2^0$ and then in ref. [12] in the case of two-body 
decays. The generality of this feature, i.e. the observability of
an edge in the $M_{l^+l^-}$ spectrum even in inclusive $l^{\pm}l^{\mp}l^{+-}$ 
and $l^+l^- + E_T^{miss}$ final states in a large part of the parameter space
was shown in [13].
We will show how much the parameters are constrained by a measurement 
of the $M_{l^+l^-}^{max}$ value of the dilepton mass spectra. Moreover, we 
will discuss a method, based on the analysis of the $M_{l^+l^-}$ spectrum, 
to find out whether the observed edge is due to the two-body or three-body 
decays of  $\tilde{\chi}_2^0$.

\section{Sparticle masses in mSUGRA} 

Within the Minimal Supersymmetric Standard Model (MSSM) the masses of
the neutralinos are determined by the parameters 
$M \left( = m_{1/2} (M_Z)\right)$,
$\mu$, and $tan\beta$ using $M_1 \simeq \frac{5}{3} tan^2{\theta}_W
\simeq 0.5 M$ ($M_1$ being the U(1) gaugino mass). In the  following,
we fix $tan\beta =2$ (we assume that $tan\beta$ could be known from previous 
experiments) and take $A_0 = 0$. In mSUGRA $\mid
\mu \mid$ quite generally turns out to be $\mid \mu \mid > M$, so that
in this case  $m(\tilde{\chi}_2^0) \simeq m(\tilde{\chi}_1^{\pm})
\simeq  2\cdot m(\tilde{\chi}_1^0) \sim M$. Both
$\tilde{\chi}_1^0$ and $\tilde{\chi}_2^0$ are gaugino like,
$\tilde{\chi}_1^0$ is almost a pure $B-ino$, and  $\tilde{\chi}_2^0$
almost a pure $W^3-ino$. In supergravity  the slepton masses are given 
by [14]: 

\begin{equation}
\hspace{2cm} m^2_{\tilde {l}_R}=m_0^2+0.15m^2_{1/2}-
                                   sin^2{\theta}_W M_Z^2cos2\beta
\end{equation}

\begin{equation}         
\hspace{2cm} m^2_{\tilde {l}_L}=m_0^2+0.52m^2_{1/2}-
                                   1/2(1-2sin^2{\theta}_W)M_Z^2cos2\beta
\end{equation}

\begin{equation}
\hspace{2cm} m^2_{\tilde {\nu}}=m_0^2+0.52m^2_{1/2}+1/2M_Z^2cos2\beta
\end{equation}

Analogous equations exist for squarks [14]. 
Therefore, when the parameters $m_0,\,m_{1/2}$ and $\tan \beta$
are known we can calculate all sparticle masses. A special case is the third
generation of squarks and sleptons, where L$-$R  mixing plays a crucial 
r\^{o}le. 

\section{Production and leptonic decay of $\tilde{\chi}_2^0$} 
Neutralinos $\tilde{\chi}_2^0$ can be produced at the LHC through a Drell-Yan
mechanism (direct production), in association with strongly interacting 
sparticles, or in the decay chain of gluinos and squarks (indirect production).
Gluino and squark pair production processes are the dominant source of 
$\tilde{\chi}_2^0$'s because of large strong interaction cross-sections. 
The branching ratios of gluino and squark decays into $\tilde{\chi}_2^0$ are 
also sizable and are shown in fig.1 as a function of the model parameters  
$m_0$ and $m_{1/2}$. One can see that the regions in ($m_0$,$m_{1/2}$) plane 
where the decays $\tilde{g} \rightarrow \tilde{\chi}_2^0 + X$ and  
$\tilde{q} \rightarrow \tilde{\chi}_2^0 + X$  are open, are 
complementary. In the region $m_0 \ga 1.47 \cdot m_{1/2}$ 
gluinos are lighter than squarks and can decay into $\tilde{\chi}_2^0$, while 
squarks prefer to decay into gluinos. In the region  
$m_0 \la 1.47 \cdot m_{1/2}$ 
squarks are lighter than gluinos and can decay into $\tilde{\chi}_2^0$, but
gluinos decay into squarks. Hence the decays  
$\tilde{g} \rightarrow \tilde{\chi}_2^0 + X$ 
($\tilde{q} \rightarrow \tilde{\chi}_2^0 + X$) and 
$\tilde{q} \rightarrow \tilde{g} \rightarrow \tilde{\chi}_2^0 + X$ 
($\tilde{g} \rightarrow \tilde{q} \rightarrow \tilde{\chi}_2^0 + X$)
can coexist (see also fig.1). 
Fig.2(a,b) shows $\sigma \times \, Br$ for indirect $\tilde{\chi}_2^0$ 
production from gluinos and squarks  as a function of $m_0$ and $m_{1/2}$.

There are three different leptonic decays of $\tilde{\chi}_2^0$ interesting 
for our study: $\tilde{\chi}_2^0 \rightarrow l^+ l^- \tilde{\chi}_1^0$,
$\tilde{\chi}_2^0 \rightarrow  \tilde{l}_R^{\pm} l^{\mp}$ and 
$\tilde{\chi}_2^0 \rightarrow  \tilde{l}_L^{\pm} l^{\mp}$. When decays
of $\tilde{\chi}_2^0$ to sleptons are allowed, sleptons
decay directly into the lightest supersymmetric particle 
(lsp$\equiv\tilde{\chi}_1^0$) with 
Br($\tilde{l}_{L,R}^{\pm}\rightarrow \tilde{\chi}_1^0 l^{\pm})=100\%$. 
Branching ratios of $\tilde{\chi}_2^0$ 
decays into leptons directly and via sleptons are shown in fig.3 as a function 
of $m_0$ and $m_{1/2}$. The regions where these decays are kinematically 
allowed are complementary in the parameter plane, depending on whether 
$\tilde{\chi}_2^0$ is lighter or heavier than $\tilde{l}_{L,R}$.
Thus, one can distinguish three domains in the ($m_0$,$m_{1/2}$) plane, 
which are (also see fig.4):

\vspace{0.5cm}

\begin{tabbing}  domain II ($0.5\cdot m_{1/2} \ga  m_0 \la 0.45\cdot
m_{1/2}\,\,\,) : \,\,\,$ 
$m_{\tilde{l}_R}<$ \= $m_{\tilde{\chi}_2^0} <  m_{\tilde{l}_L}
\,\,\,\,\,\,$ 
\= $-$ \= $\,\,\,\,\,\,\, \tilde{\chi}_2^0 \rightarrow \tilde{l}_R l$  \kill
  
domain I ($m_0 \ga 0.5\cdot m_{1/2},\,\, m_{1/2} \la 200\,{\rm GeV} \,\,\, ) : 
\,\,\,$ \> $m_{\tilde{\chi}_2^0} < m_{\tilde{l}_{L,R}}$ \>
  
 \> $\,\,\, , \,\,\,$  with 
$\tilde{\chi}_2^0 \rightarrow \tilde{\chi}_1^0\,l\,l$ \\
\\ domain II ($0.45\cdot m_{1/2} \la  m_0 \la 0.5\cdot m_{1/2} \,\,\,) : 
\,\,\,$  

$m_{\tilde{l}_R}<$ \> $m_{\tilde{\chi}_2^0} <  m_{\tilde{l}_L}$ 
\>  \> $\,\,\, , \,\,\,$  with 
$\tilde{\chi}_2^0 \rightarrow \tilde{l}_R\, l $ \\  
\\ domain III ($m_0 \la 0.45\cdot m_{1/2} \,\,\, ) : \,\,\, $
\> $m_{\tilde{\chi}_2^0} > m_{\tilde{l}_L}$  
\> \>  $\,\,\, , \,\,\,$  with   
$\tilde{\chi}_2^0 \rightarrow \tilde{l}_L\, l $ \\  
\end{tabbing}  

\vspace{0.5cm} In domain III, the decay $\tilde{\chi}_2^0
\rightarrow \tilde{l}_R l$ would also be kinematically allowed, but 
since the $B-ino$ 
component of $\tilde{\chi}_2^0$ is very small, the coupling to $\tilde{l}_R l$ 
is also small. Therefore, the decay $\tilde{\chi}_2^0 \rightarrow 
\tilde{l}_R l$ is very much suppressed in the whole domain. 

In fig.5 we show the regions for $\sigma \times Br(\tilde{\chi}_2^0
\rightarrow \tilde{\chi}_1^0 l^+ l^-) \sim 1$ and 0.1 pb,
 $\sigma \times Br(\tilde{\chi}_2^0  \rightarrow \tilde{l}_R^{\pm} l^{\mp}
\rightarrow  \tilde{\chi}_1^0 l^+ l^-) \sim 1$, 0.1 pb
and  $\sigma \times Br(\tilde{\chi}_2^0  \rightarrow \tilde{l}_L^{\pm} l^{\mp}
  \rightarrow \tilde{\chi}_1^0 l^+ l^-) \sim 1$, 
0.002 pb in the ($m_0,m_{1/2}$) plane from indirect and associated 
$\tilde{\chi}_2^0$ production followed by decays to $\tilde{\chi}_1^0 l^+l^-$
final states directly or via sleptons. 
One can see that there are regions in domains II and III  
where the mentioned decays coexist.
Finally, in fig.6 we show  $\sigma \times \, Br$ for indirect and associated
production of  $\tilde{\chi}_2^0$ decaying into leptons directly or via
sleptons. 

\section{Determination of $m_0, \, m_{1/2}$ and sparticle masses}
In order to determine the parameters $m_0$ and $m_{1/2}$ we will exploit in 
the following the kinematical features of the two- and three-body 
leptonic decays of $\tilde{\chi}_2^0$.  

As pointed out in [11], the decay 
$\tilde{\chi}_2^0 \rightarrow l^+  l^-  \tilde{\chi}_1^0$ (domain I)
has the useful kinematical property that the invariant mass of the two leptons 
$M_{l^+l^-}$  has a maximum at  

\begin{equation} 
M^{max}_{l^{+}l^{-}} = m_{\tilde{\chi}^0_2} - m_{\tilde{\chi}^0_1},
\end{equation}

whereas for the decays  $\tilde{\chi}_2^0 \rightarrow \tilde{l}_{L,R}^{\pm}
l^{\mp} \rightarrow  l^+  l^-  \tilde{\chi}_1^0 $ 
(domain II and III) the maximum of $M_{l^+l^-}$ is given by [12]:

\begin{equation} 
M^{max}_{l^{+}l^{-}} =  \frac{\sqrt{ (m^2_{\tilde{\chi}^0_2} - 
m^2_{\tilde{l}}) 
(m^2_{\tilde{l}} - m^2_{\tilde{\chi}^0_1}) }} {m_{\tilde{l}} }.
\end{equation}

Thus, the $M_{l^+l^-}$ distribution has a very characteristic shape with a  
sharp edge at the kinematical endpoint $M^{max}_{l^{+}l^{-}}$.

As the main source of $\tilde{\chi}_2^0$'s is their 
indirect production in  gluino and squark decays, 
the most suitable  signature for selecting the $\tilde{\chi}^0_2$ 
decays is provided by the topology with  
two same--flavour opposite--sign leptons accompanied by large missing 
transverse energy and usually accompanied by a high multiplicity of jets.
In this paper we thus concentrate on the
$two\,same$--$flavour,\,opposite$--$sign\,leptons + E_T^{miss}+(jets)$ channel,
where the final state leptons are electrons and muons.

\subsection{CMS detector simulation}
The simulations are done at the particle level, with parametrised detector 
responses based on detailed detector simulations. These parametrisations
are adequate for the level of detector properties we want to 
investigate, and are the only practical ones in view of the multiplicity
and complexity of the final state signal and background channels 
investigated. The essential ingredients for the investigation of SUSY 
channels are the response to jets, $E_T^{miss}$, the lepton identification
and isolation capabilities of the detector, and the capability to tag b-jets.

  The CMS detector simulation program CMSJET 3.2 [15] is used. 
It incorporates the full electro-magnetic (ECAL) and hadronic (HCAL)
calorimeter 
granularity, and includes main calorimeter system cracks in rapidity and 
azimuth. The energy resolutions for muons, electrons (photons), 
hadrons and jets are parametrised. Transverse and longitudinal 
shower profiles are also included through appropriate parametrisations.
The main detector features incorporated in the Monte--Carlo description are:

\vspace{0.2cm}
 $\bullet$ Hadronic tracks, muons and electrons  are measured up to  
$\mid\eta\mid$=2.4   

\vspace{0.15cm}

 $\bullet$ Deflection of charged particles due to the 4 T magnetic field 
           is included. 

\vspace{0.15cm}

 $\bullet$ The resolution for the muon system is parametrised according to 
[16].

\vspace{0.15cm}
 $\bullet$ The calorimetric coverage goes up to
                                  at $\mid\eta\mid$=5 for the HCAL and
                                   $\mid\eta\mid$=2.6 for the ECAL.     

\vspace{0.15cm}
 $\bullet$ ECAL energy resolution parametrized as:
 
\vspace{0.15cm}
  \hspace{4 cm} $\boldmath{ \Delta E/E=5 \% /\sqrt{E} \oplus 0.5\% }$

 $\bullet$ HCAL energy resolution is parametrised according to [17] as a
function of $\eta$; a typical hadron resolution is:  

  \hspace{4 cm} $\boldmath{ \Delta E/E=80 \% /\sqrt{E} \oplus 7\% }$ 
\vspace{0.15cm}

 $\bullet$ Energy resolution for very forward calorimeter (VFCAL), in the
parallel plate chambers option:

  \hspace{4 cm} $\boldmath{ \Delta E/E=90\% /\sqrt{E} \oplus 3\% }$




\vspace{0.15cm}
 $\bullet$ Granularity of calorimeters:

\vspace{0.2cm}

\begin{center}
\begin{tabular}{|c|c|c|} 
\hline 
 & $\eta-range$ & $\Delta \varphi$ x $\Delta \eta$  \\
\hline
\hline
\hline
ECAL(barrel) & $\mid\eta\mid<1.57$     &0.015x0.015 \\
             &$1.57<\mid\eta\mid<1.65$ & crack  \\
\hline
ECAL(endcap) &$1.65<\mid\eta\mid<2$   &0.022x0.022 \\
             &$2.<\mid\eta\mid<2.35$  &0.029x0.029 \\
             &$2.35<\mid\eta\mid<2.61$&0.043x0.043 \\
\hline  
\hline
HCAL & $\mid\eta\mid<2.26$      & 0.087x0.087 \\
     & $2.26<\mid\eta\mid<2.6$  & 0.174x0.175 \\
     & $2.6<\mid\eta\mid<3$     & 0.195x0.349   \\
\hline  
\hline
VFCAL& $3<\mid\eta\mid<4$ &10x10 $cm^2$  \\
    &  $4<\mid\eta\mid<5$ & 5x5 $cm^2$ \\        
\hline
\end{tabular}
\end{center}

\vspace{0.2cm}

 $\bullet$ $E_{threshold}$ on cells:

 \hspace{4 cm} - ECAL: $E_{threshold}$ =50 MeV
\vspace{0.15cm}

 \hspace{4 cm} - HCAL: $E_{threshold}$ =250 MeV
\vspace{0.15cm}

 \hspace{4 cm} - VFCAL: $E_{threshold}$ =500 MeV
\vspace{0.15cm}

 $\bullet$ a modified UA1- jet finding algorithm with a cone size of
 $\Delta R$=0.9 (for description see CMSJET 3.5 [15]) is used for jet 
 reconstruction. \\

\subsection{Observability of edges in invariant dilepton mass distributions}

In this chapter we determine the regions in the ($m_0,\,m_{1/2}$) parameter
plane, where the characteristic edge
in the $M_{l^+l^-}$ distribution can be observed in inclusive final states with
$two\,same$--$flavour, \,opposite$--$sign\,leptons + E_T^{miss}+(jets)$ 
with different luminosities at LHC.

Standard Model background processes are generated with PYTHIA 5.7 [18].
We use CTEQ2L structure functions.
The largest  background is due to $t\bar{t}$ production,
with both $W$'s decaying into leptons, or one of the leptons  from a $W$
decay and the other from the $b$-decay of the same $t$-quark.  We also
considered other SM backgrounds: $W+jets$, WW, WZ, $b \bar b$
and  $\tau \tau$-pair production, with decays into electrons and muons.
Chargino pair production $\tilde{\chi}_1^{\pm}\tilde{\chi}_1^{\mp}$ is 
the largest SUSY background  but gives a small contribution compared
to the signal. 

To observe an  
edge in the $M_{l^+l^-}$ distributions with the statistics provided
by an integrated luminosity  $L_{int}=10^3 \,pb^{-1}$ in a significant part
of the ($m_0,m_{1/2}$) parameter plane, it is enough to require two 
hard isolated leptons ($p_T^{l_{1,2}}>15$ GeV) accompanied by large missing 
energy, $E_T^{miss}>100$ GeV. Our criterion for observing an edge in the 
$M_{l^+l^-}$ distribution contains two requirements:
$(N_{EV}-N_B)/\sqrt{N_{EV}}\ga 5$ and $(N_{EV}-N_B)/N_B \ga 1.3$, where 
$N_{EV}$ is the number of events with $M_{l^+l^-} \le M_{l^+l^-}^{max}$, 
and $N_B$ is number of the expected background events.  
Fig.7 shows the invariant mass spectra of the two leptons at various
($m_0,m_{1/2}$) points from domains I, II and III, respectively. 
The observability  of the "edge"
varies from $77 \sigma$ and signal to background ratio
31 at point (200,160) to $27  \sigma$ and a signal to background ratio 2.3 at 
point (60,230).
The appearance of the edges in the distributions is sufficiently pronounced 
already with $L_{int}=10^3$ pb$^{-1}$ in a significant part of 
$(m_0,m_{1/2})$ parameter plane, see fig.9. 
The edge position can be measured with a precision of $\sim 0.5$ GeV.

With increasing  $m_0$ and $m_{1/2}$ cross-sections  are 
decreasing, therefore higher luminosity and harder cuts 
are needed. To achieve maximal reach in $m_{1/2}$ with 
$L_{int}=10^4 \,pb^{-1}$ for points from domain III, a cut up to  
$E_T^{miss}>300$ GeV is necessary to suppress the background sufficiently.  
For points with large $m_0$ (domain I) the transverse momentum $p_T$ of the 
leptons and $E_T^{miss}$ are not very large, but there are more 
hard jets due to gluino and squark decays.
Thus for these points we keep the same cuts for leptons and missing
energy as before ($p_T^{l_{1,2}}>15$ GeV, $E_T^{miss}>100$ GeV) 
and require in addition a jet multiplicity $N_{jet}\ge 3$, 
with energy $E_T^{jet}>100$ GeV, in the rapidity range 
$\mid\eta_{jet}\mid<3.5$. To optimise the edge visibility 
we also apply an azimuthal angle cut, 
$\Delta \phi(l^+l^-) < 120^0$.  For points from domain II, the jet 
multiplicity requirement is also helpful. Right sleptons are too light to 
provide large lepton $p_T$ and $E_T^{miss}$, and to use 
cuts on $p_T^l$ and $E_T^{miss}$ alone is not very advantageous.
With $L_{int}=10^5 \,pb^{-1}$, to suppress the background  at larger 
accessible $m_0,\,m_{1/2}$ values, 
we have to require at least 2 or 3 jets, depending on the $m_0,\,m_{1/2}$ 
region to be explored.
Fig.8 shows invariant dilepton mass distributions at some 
($m_0,m_{1/2}$) points close to maximum reach with
$L_{int}=10^4 \,pb^{-1}$ and $L_{int}=10^5 \,pb^{-1}$ respectively.  

The regions of the ($m_0,m_{1/2}$) parameter plane where an edge in the 
$M_{l^+l^-}$ spectra can 
be observed at different luminosities are shown in fig.9. In fig.10 we show 
separately the three domains where an edge due to $\tilde{\chi}_2^0 
\rightarrow l l \tilde{\chi}_1^0$, $\tilde{l}_R l$ and $\tilde{l}_L l$
decays can be observed at $L_{int}=10^3 \,pb^{-1}$.
One can notice a small overlapping region, where we expect to observe 
two edges, due to $\tilde{\chi}_2^0 \rightarrow l^+ l^- \tilde{\chi}_1^0$ 
and to $\tilde{\chi}_2^0 \rightarrow \tilde{l}_R^{\pm} l^{\mp} 
\rightarrow l^+ l^- \tilde{\chi}_1^0$
decays ({\it case 1}). With increasing luminosity 
and correspondingly higher statistics, this overlapping region 
increases, see figs.11 and 12. These plots 
show the same as fig.10, but for $L_{int}=10^4 \,pb^{-1}$ and
$L_{int}=10^5 \,pb^{-1}$, respectively. 
An additional region appears where two edges can be observed simultaneously, 
due to $\tilde{\chi}_2^0 \rightarrow \tilde{l}_R^{\pm} l^{\mp}
\rightarrow l^+ l^- \tilde{\chi}_1^0$ and
$\tilde{\chi}_2^0 \rightarrow \tilde{l}_L^{\pm} l^{\mp}
\rightarrow l^+ l^- \tilde{\chi}_1^0$ decays ({\it case 2}). These
regions ({\it case 1} and {\it 2}) are due to the
coexistence of different $\tilde{\chi}_2^0$ decay modes has been 
seen in fig.5.
An example of a $M_{l^+l^-}$ distribution for {\it case 1} is shown
in fig.13.

Therefore, to a given integrated luminosity at LHC ($L_{int}=10^3$ pb$^{-1}$ 
to $10^5$ pb$^{-1}$) there corresponds a definite parameter region where 
the characteristic structure in the $M_{l^+l^-}$ distribution can be seen.
This fact already gives a preliminary information 
about the parameters $m_0$ and $m_{1/2}$. The observation of two edges would 
give even stronger constraints.

\subsection{$M_{l^+l^-}^{max}$ analysis of ($m_0,m_{1/2}$) parameter plane}

Within mSUGRA all sparticle masses for every point ($m_0,m_{1/2}$) can be
calculated. The expected edge position $M_{l^+l^-}^{max}$ in the dilepton mass 
distribution can  then  be obtained from eqs.(4) and (5).
In fig.14 we show contours for various expected 
values of $M_{l^+l^-}^{max}$ in the ($m_0,m_{1/2}$) parameter plane. 
Different lines with the same value of $M_{l^+l^-}^{max}$ belong to domains
I, II and III (with the corresponding decay mode of $\tilde{\chi}_2^0$).
The region of $M_{l^+l^-}^{max}$ becoming accessible at LHC is:

\begin{tabbing} 
for 
$\hspace{2cm}$ \= $\tilde{\chi}_2^0 \rightarrow \tilde{\chi}_1^0  l^+ l^- 
\hspace{1cm}$ \= $- \hspace{1cm}$ 50 GeV $\la$ \= $M_{l^+l^-}^{max} \la$ 
90 GeV\\
\\
\> $\tilde{\chi}_2^0 \rightarrow \tilde{l}_R l$ \> $-$ \>
$M_{l^+l^-}^{max} \ga 10$ GeV \\
\\
\> $\tilde{\chi}_2^0 \rightarrow \tilde{l}_L l$ \> $-$ \>
$M_{l^+l^-}^{max} \ga 20$ GeV \\ 
\end{tabbing}

More specifically,
at low luminosity, $L_{int}=10^3$ pb $^{-1}$, at the beginning of the LHC 
operation, the accessible values of $M_{l^+l^-}^{max}$ 
lie in the following ranges (see figs.10 and 14):

\begin{tabbing} 
for
$\hspace{2cm}$ \= $\tilde{\chi}_2^0 \rightarrow \tilde{\chi}_1^0  l^+ l^- 
\hspace{1cm}$ \= $- \hspace{1cm}$ 50 GeV $\la$ \= $M_{l^+l^-}^{max} \la$ 
80 GeV \\
\\
\> $\tilde{\chi}_2^0 \rightarrow \tilde{l}_R l$ \> $-\hspace{1cm}$ 10 GeV $\la$
\> $M_{l^+l^-}^{max} \la 110 $ GeV \\
\\
\> $\tilde{\chi}_2^0 \rightarrow \tilde{l}_L l$ \> $- \hspace{1cm}$ 20 GeV 
$\la$ \> $M_{l^+l^-}^{max} \la 120$ GeV \\ 
\end{tabbing}

It follows from the discussion above that a measurement of 
$M_{l^+l^-}^{max}$ in the dilepton mass distribution, with a single edge, 
constrains the parameters in general to three lines in the ($m_0,m_{1/2}$) 
parameter plane. In case of $M_{l^+l^-}^{max} \ga 90$ GeV the constraint is 
stronger, there are just two possible lines. The most favourable case is when
the measured $M_{l^+l^-}^{max}$ value is large, $M_{l^+l^-}^{max} \ga 180$ GeV.
Then one is left with a single line in the ($m_0,m_{1/2}$) parameter plane.
For the present study we have chosen, as an example of the general situation,
the case of $M_{l^+l^-}^{max}=74 \pm 1$ GeV, with three lines 
corresponding to the domains I,II and III, respectively.
The next step is to find out which line in the ($m_0,m_{1/2}$) plane is 
the right one. To this purpose we
have analysed points along these lines, given in tables 1-3. 
 
The study is made for the low luminosity case, 
$L_{int}=10^3$ pb$^{-1}$.

\vspace{0.2cm} 
\begin{center}
Table 1. $M_{l^+l^-}^{max}$ values (in GeV) at the investigated 
$(m_0,m_{1/2})$ points from domain I,
$\tilde{\chi}_2^0 \rightarrow \tilde{\chi}_1^0 + l^+ + l^-$.
\end{center}

\begin{center}
\begin{tabular}{| c |
@{\extracolsep{0.5mm}}c@{\extracolsep{2mm}}|
@{\extracolsep{0.5mm}}c@{\extracolsep{2mm}}|
@{\extracolsep{0.5mm}}c@{\extracolsep{2mm}}|
@{\extracolsep{0.5mm}}c@{\extracolsep{2mm}}|
@{\extracolsep{0.5mm}}c@{\extracolsep{2mm}}|
@{\extracolsep{0.5mm}}c@{\extracolsep{2mm}}|
@{\extracolsep{0.5mm}}c@{\extracolsep{2mm}}|
@{\extracolsep{0.5mm}}c@{\extracolsep{2mm}}|}
\hline 
&(120,160)&(130,160)&(180,160)&(200,160)&(220,160)&(240,160)&
(290,160)&(350,160)\\
\hline
\hline
\hline
$M_{l^+l^-}^{max}$ & 74 & 74 & 74 & 74 & 74& 74& 74&  73.7\\
\hline
\end{tabular}
\end{center}

\vspace{0.2cm} 

\begin{center}
Table 2. $M_{l^+l^-}^{max}$ values (in GeV) at the investigated 
$(m_0,m_{1/2})$ points from domain II,
$\tilde{\chi}_2^0 \rightarrow \tilde{l}_R^{\pm} + l^{\mp}
\rightarrow \tilde{\chi}_1^0 + l^+ + l^-$.
\end{center}

\begin{center} 
\begin{tabular}{|c|c|c|c|c|c|} 
\hline 
$(m_0,m_{1/2})\rightarrow$ &(80,162)&(90,170)&(105,180)&(110,187)&(120,195) \\
\hline
\hline
\hline 
$M_{l^+l^-}^{max}$ & 74 & 74 & 73 & 75 & 73 \\
\hline
\end{tabular}
\end{center} 

\vspace{0.2cm}
\begin{center}
Table 3. $M_{l^+l^-}^{max}$ values (in GeV) at the investigated 
$(m_0,m_{1/2})$ points from domain III,
$\tilde{\chi}_2^0 \rightarrow \tilde{l}_L^{\pm} + l^{\mp} 
\rightarrow \tilde{\chi}_1^0 + l^+ + l^-$.
\end{center}

\begin{center} 
\begin{tabular}{|c|c|c|c|c|} 
\hline 
$(m_0,m_{1/2})\rightarrow$  &(20,195)&(40,210)&(60,230)&(80,255) \\
\hline
\hline
\hline
$M_{l^+l^-}^{max}$ & 73 & 73 & 73 &  73 \\
\hline
\end{tabular}
\end{center} 

One should first
notice that the observation of two edges at $L_{int}=10^3$ pb$^{-1}$
would determine the ($m_0,m_{1/2}$) point uniquely. This is due to the fact 
that the set of the edge position values in the $M_{l^+l^-}$ spectrum is 
different at each point of the parameter region ({\it case 1}), 
where two edges are expected to be observed at 
$10^3$ pb$^{-1}$, see figs.10 and 14. With a luminosity $L_{int}=10^4$ 
pb$^{-1}$ the positions of the two edges will fix two ($m_0,m_{1/2}$) 
points, belonging to  
domains II and III and corresponding to {\it case 1} and 
{\it case 2}, respectively.
At high luminosity, $L_{int}=10^5$ pb$^{-1}$, the observation of two edges
can give up to three possible ($m_0,m_{1/2}$) points. One of them is from 
domain II, corresponding  to {\it case 1}. The lines $M_{l^+l^-}^{max}=const$ 
corresponding to 
$\tilde{\chi}_2^0\rightarrow \tilde{l}_R l$ decays have the form of an 
ellipse and can cross the $M_{l^+l^-}^{max}=const$ lines corresponding to  
$\tilde{\chi}_2^0\rightarrow \tilde{l}_L l$ decays twice. Hence, two points 
with the same set of edge positions in the $M_{l^+l^-}$ spectrum can be found
in domain III corresponding to {\it case 2}.
A discrimination between these points is possible on basis of the event 
kinematics, and/or by an analysis of the total event rate and the relative 
number of events corresponding to the two peaks, (see figs.5,10-12).

\subsection{Discrimination between different $\tilde{\chi}_2^0$ leptonic 
decays} 

For a definite value of the edge position $M_{l^+l^-}^{max}$ one expects a 
different shape of the $M_{l^+l^-}$ distributions in two- and three-body 
decays (see fig.13, where the first peak is due to a two-body decay 
of $\tilde{\chi}_2^0$ and the second one due to a three-body decay). 
As we have seen from figs.7 and 8 the signal events contribute 
in the interval $0\la M_{l^+l^-} \la M_{l^+l^-}^{max}$. In the following we 
only consider events in this mass region. The average value $<$$M_{l^+l^-}$$>$ 
of signal and background events of this mass region is shown in fig.15
as a function of $m_0$. The errors are calculated by taking into account the 
statistical error, a systematic error in the measurement of the edge position, 
and a systematic error of 30 $\%$ for background uncertainty (the main 
background is $t\bar t$). One clearly sees that $<$$M_{l^+l^-}$$>$ is 
significantly smaller in the case of a direct three-body decay 
$\tilde{\chi}_2^0 \rightarrow l l \tilde{\chi}_1^0$. Thus the shape of the 
dilepton mass spectrum already allows one to decide whether $\tilde{\chi}_2^0$
decays into a slepton or not. 

In order to distinguish between domains II and III, we suggest to use 
the fact that in general the contour lines with the same $M_{l^+l^-}^{max}$ 
for right and left sleptons  have no overlap in $m_{1/2}$ in regions of 
parameter space which are 
accessible at a given luminosity, see figs.9 and 14
as an example. It means that the masses of $\tilde{\chi}_1^0$'s are different 
for these two lines, and hence $E_T^{miss}$ is expected to be 
different. In fig.16 we show the $<$$E_T^{miss}$$>$ values for  events with
$\tilde{\chi}_2^0 \rightarrow \tilde{l}_{L,R}^{\pm}l^{\mp} \rightarrow l^+ l^- 
\tilde{\chi}_1^0$ decays after the cuts $p_T^{l_{1,2}}>15$ GeV and 
$E_T^{miss}>100$ GeV, $M_{l^+l^-}<M_{l^+l^-}^{max}$. 
The errors are calculated by taking 
into account the statistical error and a systematic error in the measurement 
of the edge position. As can be seen from fig.16, $\,<$$E_T^{miss}$$>\,$ is 
larger in the case of
$\tilde{\chi}_2^0 \rightarrow \tilde{l}_L^{\pm} l^{\mp}\rightarrow
l^+ l^- \tilde{\chi}_1^0$ than in the case of
$\tilde{\chi}_2^0 \rightarrow \tilde{l}_R^{\pm} l^{\mp} 
 \rightarrow l^+ l^- \tilde{\chi}_1^0$ as expected.

\subsubsection{Event rate analysis} 
When the correct $M_{l^+l^-}^{max}$ line is chosen, the last step is to find 
the point ($m_0,m_{1/2}$) on this line. In general the cross section 
falls with increasing $m_{1/2}$ and $m_0$. Thus, we study the event 
rate along the corresponding $M_{l^+l^-}^{max}$ line. We first discuss the 
domain III, where the situation is simpler. 
For the event rate analysis at $L_{int}=10^3$ pb$^{-1}$,
to reduce the uncertainties due to background,
we use a harder cut on $E_T^{miss}$, $E_T^{miss}>130$ GeV. 
The dependence of the expected 
event rate on $m_0$ is shown in fig.17. The errors are calculated by taking 
into account the statistical error and a systematic error of 30 $\%$ for 
background uncertainty. A systematic 
error due to the precision of the edge position measurement is also taken
into account. From the observed event rate 
we can then determine $m_0$ with a good accuracy, $\delta m_0 \simeq 4$ GeV. 
The parameter $m_{1/2}$ is then given by the $M_{l^+l^-}^{max}-$line in the 
$(m_0,m_{1/2})$ plane. The precision obtained in such a way 
is $\delta m_{1/2} \simeq 4$  GeV.

In domain II, the event rate along a line of definite $M_{l^+l^-}^{max}$ is 
first increasing and then decreasing with $m_0$, see fig.18. This is mainly 
due to the change in the branching ratios (see fig.4). 
The dependence of the event rate is, however, such that 
$m_0$ cannot be determined uniquely. To a given event rate there correspond 
in general two $m_0$ values. The ambiguity can, however, 
be solved at  high luminosity $L_{int}=10^5$ pb$^{-1}$, when  
two edges in the $M_{l^+l^-}$ distribution can be observed. 

For domain I, the $m_0$ dependence of the event rate is shown in fig.19a, 
again for $M_{l^+l^-}^{max}\simeq 74 \pm 1$ GeV. Notice the steep increase of 
the rate at $m_0 \simeq 120-130$ GeV. This is due to the fact that the decay  
channel $\tilde{\chi}_2^0 \rightarrow l l \tilde{\chi}_1^0$ is just opening in 
this region. As can be seen from the curve in fig.19a, there is an ambiguity 
in the determination of $m_0$ if the event rate is in the region $3700\la 
N_{EV} \la 5600$  or 120 GeV $\la m_0 \la $ 240 GeV. Here it helps if we look 
at the average number of jets $<$$N_{jet}$$>$ in the events under study.
Fig.19b shows $<$$N_{jet}$$>$ as a function of $m_0$. $<$$N_{jet}$$>$ is 
rising with $m_0$ as more jets are produced as the squarks become heavier. 
With the measured $<$$N_{jet}$$>$ we can resolve the ambiguity in the 
mentioned region 120 GeV $\la m_0 \la $ 240 GeV and thus determine $m_0$
with $\delta m_0 \simeq 7-3$ GeV.     

\section{Conclusions}

In this paper we have performed a detailed analysis of events with the 
signature $l^+l^- + E_T^{miss} + (jets)$ to be expected in pp collisions at
LHC. Our aim has been to determine the parameters $m_0$ and $m_{1/2}$ 
of the Minimal Supergravity Model and 
to get information on the mass spectrum of SUSY particles, 
assuming knowledge of $tan \beta$ from previous experiments.
 We have exploited the
property of the $\tilde{\chi}_2^0$ leptonic decays $\tilde{\chi}_2^0
\rightarrow l^+ l^-  \tilde{\chi}_1^0$,  $\tilde{\chi}_2^0 \rightarrow   
\tilde{l}_{L,R} l \rightarrow l^+ l^-  \tilde{\chi}_1^0$ 
that the invariant mass of the two final leptons has a 
maximum,  $M_{l^+l^-}^{max}$, clearly visible even in inclusive production. 
We have determined for different luminosities the regions in the 
($m_0,m_{1/2}$) parameter plane where 
one or two edges can be observed in the invariant dilepton mass distributions.
These regions already give preliminary information
about the model parameters.
The appearance of the edges in the $M_{l^+l^-}$ distributions can be 
already seen with a luminosity $L_{int}=10^3$ pb$^{-1}$. Therefore 
we have concentrated on a low
luminosity study. On the other hand, in case no such observation will be 
made at this luminosity, the corresponding parameter region can be 
excluded, and the same analysis can be done at  higher luminosity.

We have shown that a measurement of the $M_{l^+l^-}^{max}$ value constrains 
the parameters mainly to three lines in the ($m_0,m_{1/2}$) parameter plane. 
The lines correspond to the decay modes 
$\tilde{\chi}_2^0 \rightarrow l^+ l^-  \tilde{\chi}_1^0$, 
$\tilde{l}_L l \rightarrow l^+ l^-  \tilde{\chi}_1^0$,
$\tilde{l}_R l \rightarrow l^+ l^-  \tilde{\chi}_1^0$ respectively. 
We have worked out a method to discriminate the
three-body from the two-body $\tilde{\chi}_2^0$ decays. In the case of 
three-body
$\tilde{\chi}_2^0$ decays the parameter $m_{1/2}$ can be determined by the 
measured value of $M_{l^+l^-}^{max}$ with a precision of $\sim 0.5$ GeV. 
The parameter $m_0$ can then be determined from the observed
event rate with a precision of 7-3 GeV. In the case of 
two-body $\tilde{\chi}_2^0$  decays, a measurement of the missing transverse 
energy can allow one to distinguish between the two possible decays
 $\tilde{\chi}_2^0 \rightarrow  \tilde{l}_L l$  and 
$\tilde{\chi}_2^0 \rightarrow  \tilde{l}_R l$, but a more detailed study is
needed. By an event rate analysis along 
the corresponding line in the ($m_0,m_{1/2}$) plane we can determine $m_0$ and
$m_{1/2}$, $\delta m_0 \sim \delta m_{1/2} \sim 4$ GeV. 

Knowing $m_0$, $m_{1/2}$ and $tan \beta$, the masses of all SUSY 
particles (except for the 3rd generation of squarks and sleptons) are 
calculated by RGE. The precisions which can be achieved are $\sim 1 - 6$ GeV. 

In such a way it is possible to obtain information about SUSY particle masses
already with low luminosity ($L = 10^3$ pb$^{-1}$) even without having direct 
experimental evidence for their existence. This is especially important for 
sleptons in a parameter region where high luminosity would be necessary to 
detect them through direct production.

This study has been performed for $tan \beta=2$, but it is  also possible
for high values of $tan \beta$. Most likely, for large $tan \beta \ga 30$ a
higher luminosity will be needed because of smaller branching ratios
of the $\tilde{\chi}_2^0$ leptonic decays.

Let us mention some further  interesting aspects of this work. Selecting
the two-body  $\tilde{\chi}_2^0$ leptonic decays 
by our method represents an indirect evidence for sleptons in the framework 
of mSUGRA. In this way it is possible to probe slepton masses
up to $\sim 740$ GeV well beyond what is possible in direct [5,6] searches.
As it has been shown in this study, the edge in the invariant dilepton 
mass distributions is expected to appear at $M_{l^+l^-}^{max}\ga 10$ GeV, 
being  quite generally a signal for a two- or three-body decay of some 
abundantly produced heavy object. Hence such an observation may serve as a 
first evidence for physics Beyond the Standard Model, and if observed with 
significant $E_T^{miss}$ it would be a clear evidence for SUSY, more 
specifically for
$\tilde{\chi}_2^0$ production.

\vspace{0.5cm}
\hspace{-0.5cm}{\Large \bf Acknowledgements}

We especially thank Howard Baer for answering many questions and for 
very useful discussions.
We would like to thank Alfred Bartl for his interest in this work and 
valuable discussions. 
L.R. thanks for financial support by the Austrian Academy of Sciences.


\clearpage
\begin{figure}
\begin{center}
   \epsfig{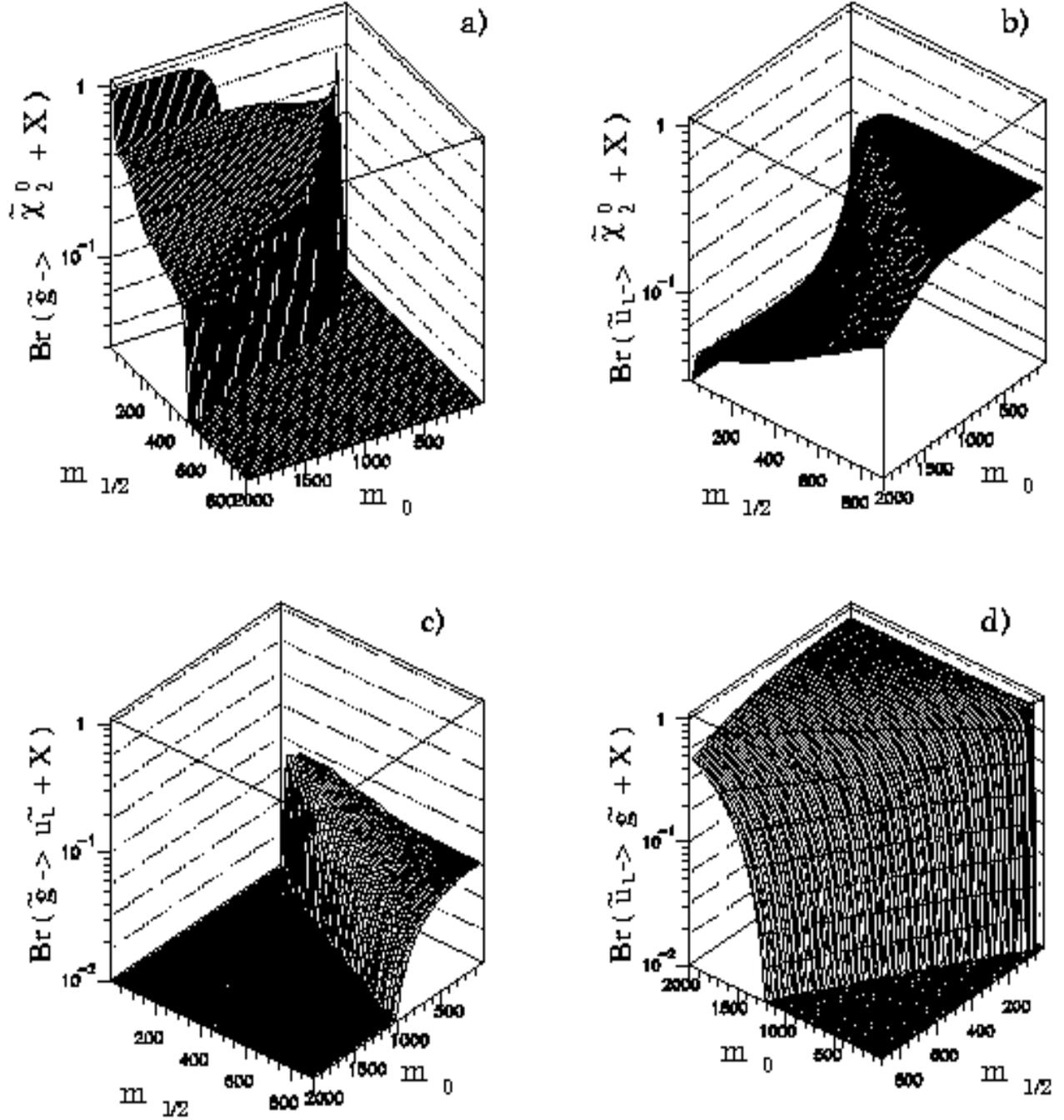}
\vspace{0.5cm}
\caption{Decay branching ratios as a function of  $m_0$ and $m_{1/2}$ (in GeV)
for: 
a) $\tilde{g} \rightarrow 
\tilde{\chi}_2^0 + X$, 
b) $\tilde{u}_L \rightarrow \tilde{\chi}_2^0 + X$ and 
c) $\tilde{g} \rightarrow \tilde{u}_L + X$,   
d) $\tilde{u}_L \rightarrow \tilde{g} + X$, for $\tan \beta = 2, \, A_0=0,
\mu < 0$.} 
\end{center}
\end{figure}

\clearpage
\begin{figure}
\begin{center}
   \epsfig{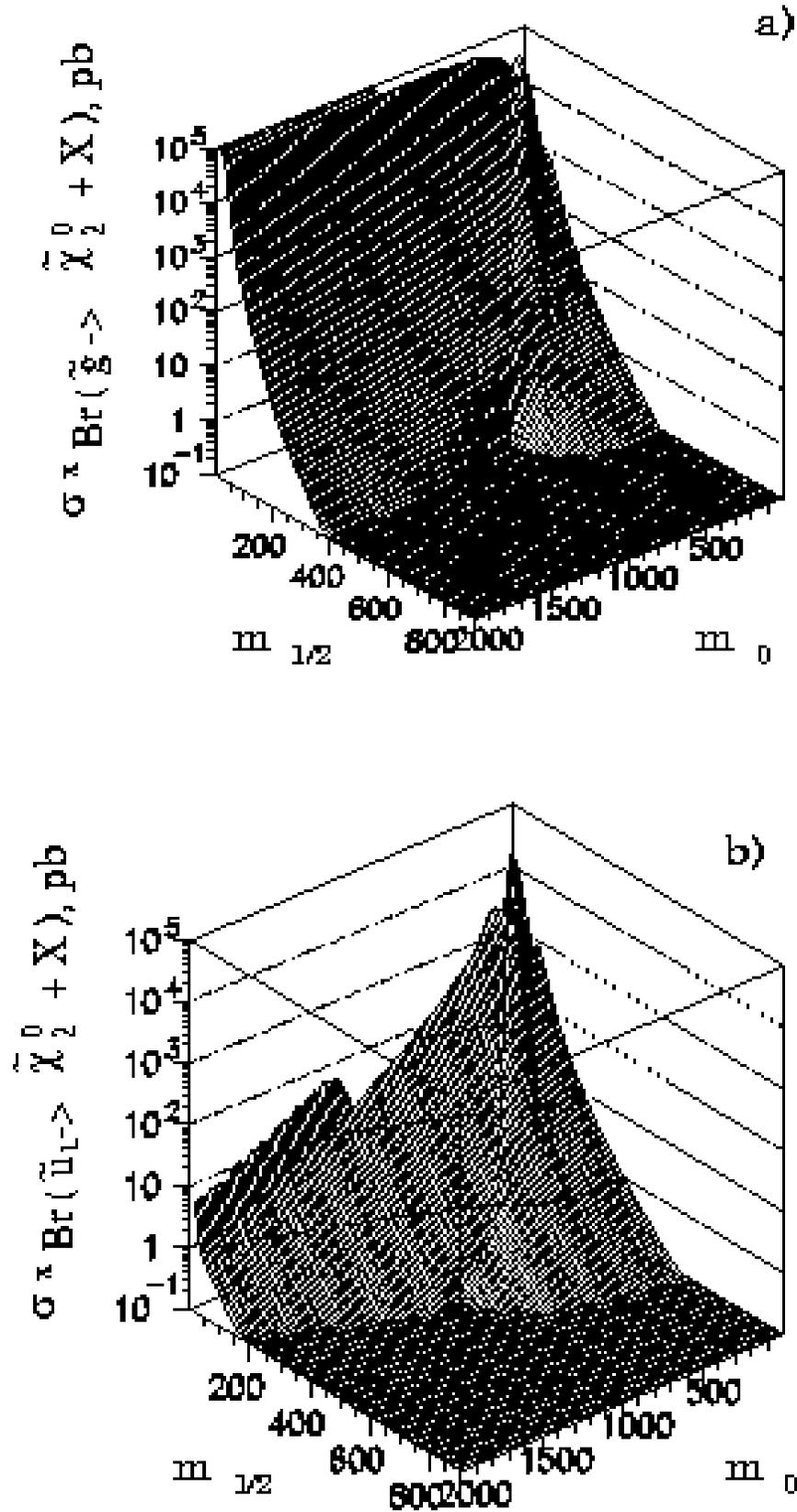}
\vspace{0.5cm}
\caption{Sigma times branching   
ratios as a function of  $m_0$ and $m_{1/2}$ (in GeV) for indirect 
$\tilde{\chi}_2^0$ production
from gluinos (a) and squarks (b), for $\tan \beta = 2, \, A_0=0, \mu < 0$.}
\end{center}
\end{figure}

\clearpage
\begin{figure}
\begin{center}
   \epsfig{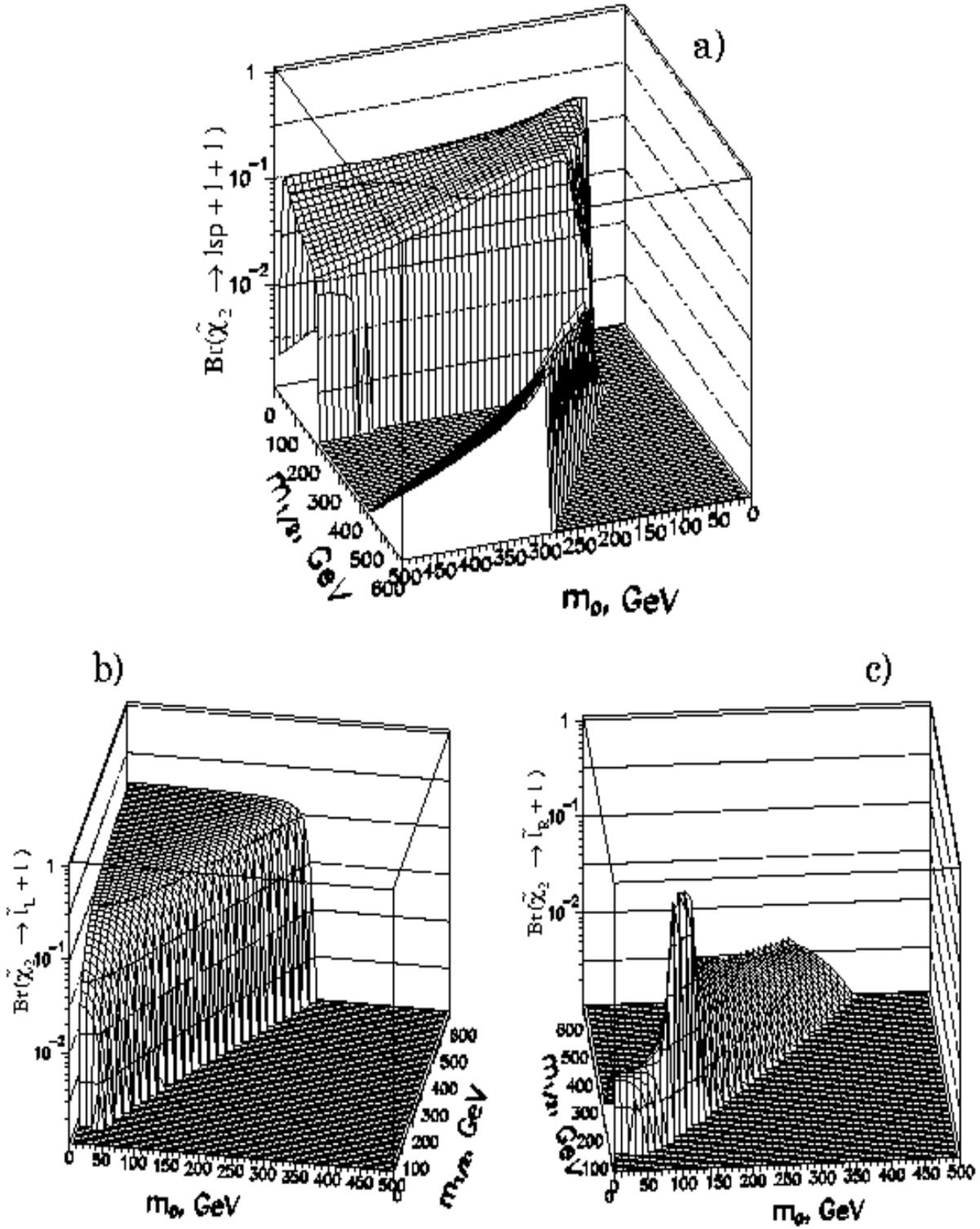}
\vspace{0.5cm}
\caption{Branching ratios of $\tilde{\chi}_2^0$ decays:
a) $\tilde{\chi}_2^0 \rightarrow \tilde{\chi}_1^0 l^+ l^-$,
b) $\tilde{\chi}_2^0 \rightarrow  \tilde{l}_L^{\pm} l^{\mp}$ and
c) $\tilde{\chi}_2^0 \rightarrow  \tilde{l}_R^{\pm} l^{\mp}$
as a function of $m_0$ and $m_{1/2}$, for $\tan \beta = 2, \, A_0=0, \mu < 0$.}
\end{center}
\end{figure}

\clearpage
\begin{figure}
\begin{center}
   \epsfig{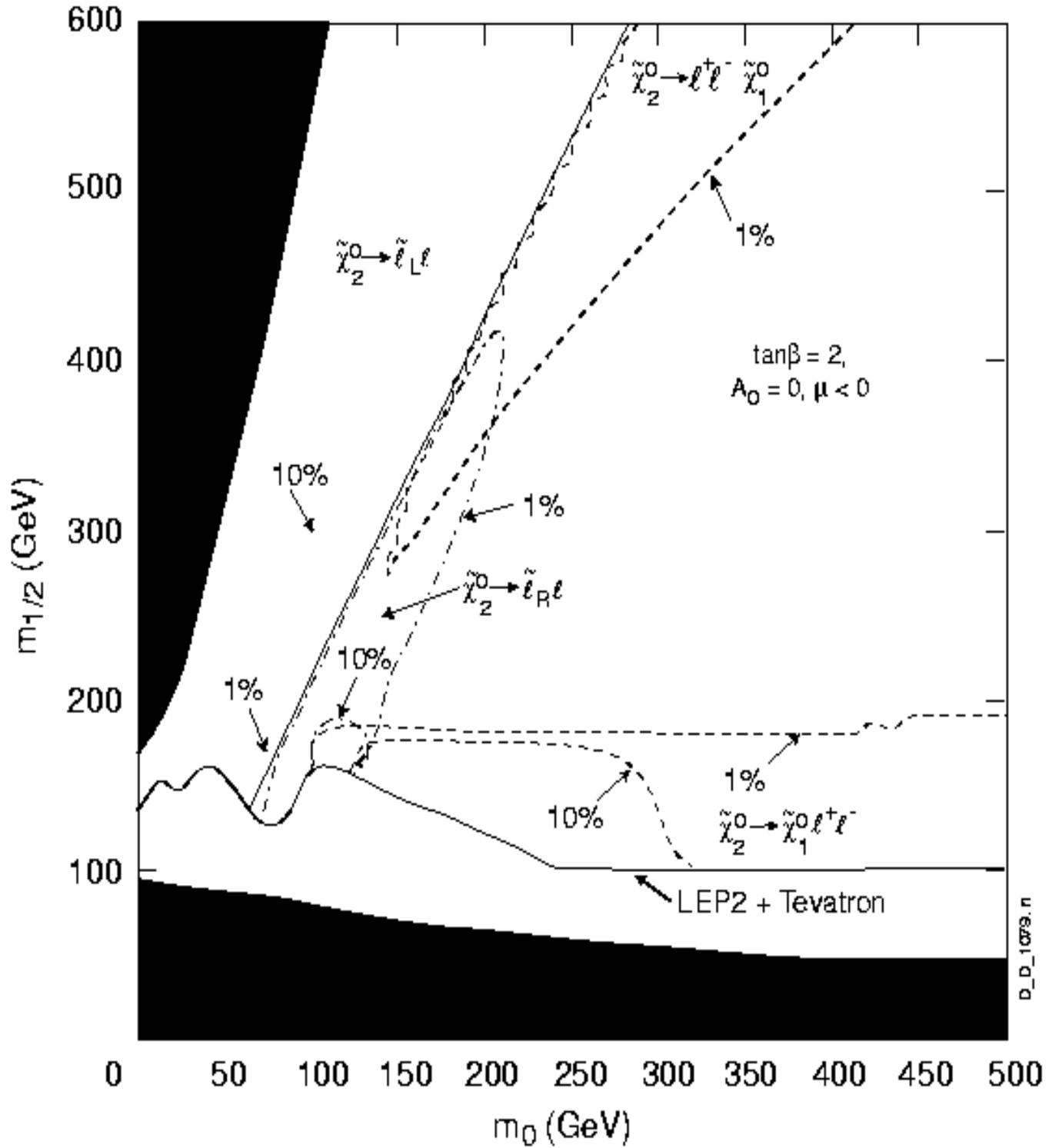}
\vspace{0.5cm}
\caption{Domains of the decays  
$\tilde{\chi}_2^0 \rightarrow \tilde{\chi}_1^0 l^+ l^-$ (dashed line),
$\tilde{\chi}_2^0 \rightarrow  \tilde{l}_L^{\pm} l^{\mp}$ (solid line) and 
$\tilde{\chi}_2^0 \rightarrow  \tilde{l}_R^{\pm} l^{\mp}$ (dashed-dotted line)
in the ($m_0,m_{1/2}$) plane, corresponding to decay branching ratios in 
excess of $1 \%$ and $10 \%$ respectively, $\tan \beta = 2, \, A_0=0, \, 
\mu < 0$.}
\end{center}
\end{figure}

\clearpage
\begin{figure}
\begin{center}
   \epsfig{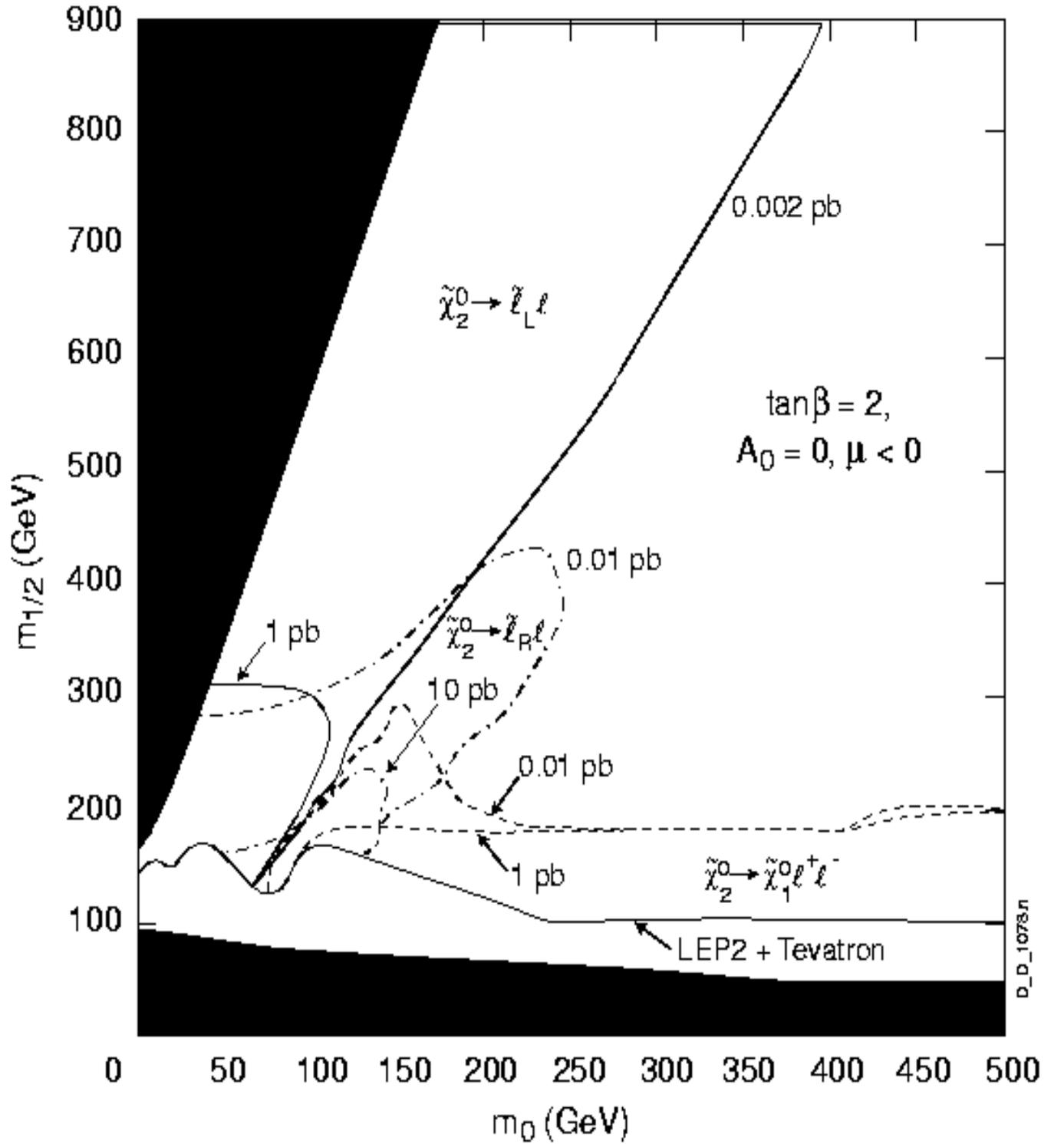}
\vspace{0.5cm}
\caption{Contour lines for cross-section times branching   
ratios in the ($m_0$, $m_{1/2}$) plane
for indirect and associated $\tilde{\chi}_2^0$ production
followed by decays:
$\tilde{\chi}_2^0 \rightarrow \tilde{\chi}_1^0 l^+ l^-$ (dashed line),
$\tilde{\chi}_2^0 \rightarrow  \tilde{l}_L^{\pm} l^{\mp}
\rightarrow \tilde{\chi}_1^0 l^+ l^-$ (solid line) and 
$\tilde{\chi}_2^0 \rightarrow  \tilde{l}_R^{\pm} l^{\mp}
\rightarrow \tilde{\chi}_1^0 l^+ l^-$ (dashed-dotted line),
$\tan \beta = 2, \, A_0=0, \, \mu < 0$.}
\end{center}
\end{figure}

\clearpage
\begin{figure}
\begin{center}
   \epsfig{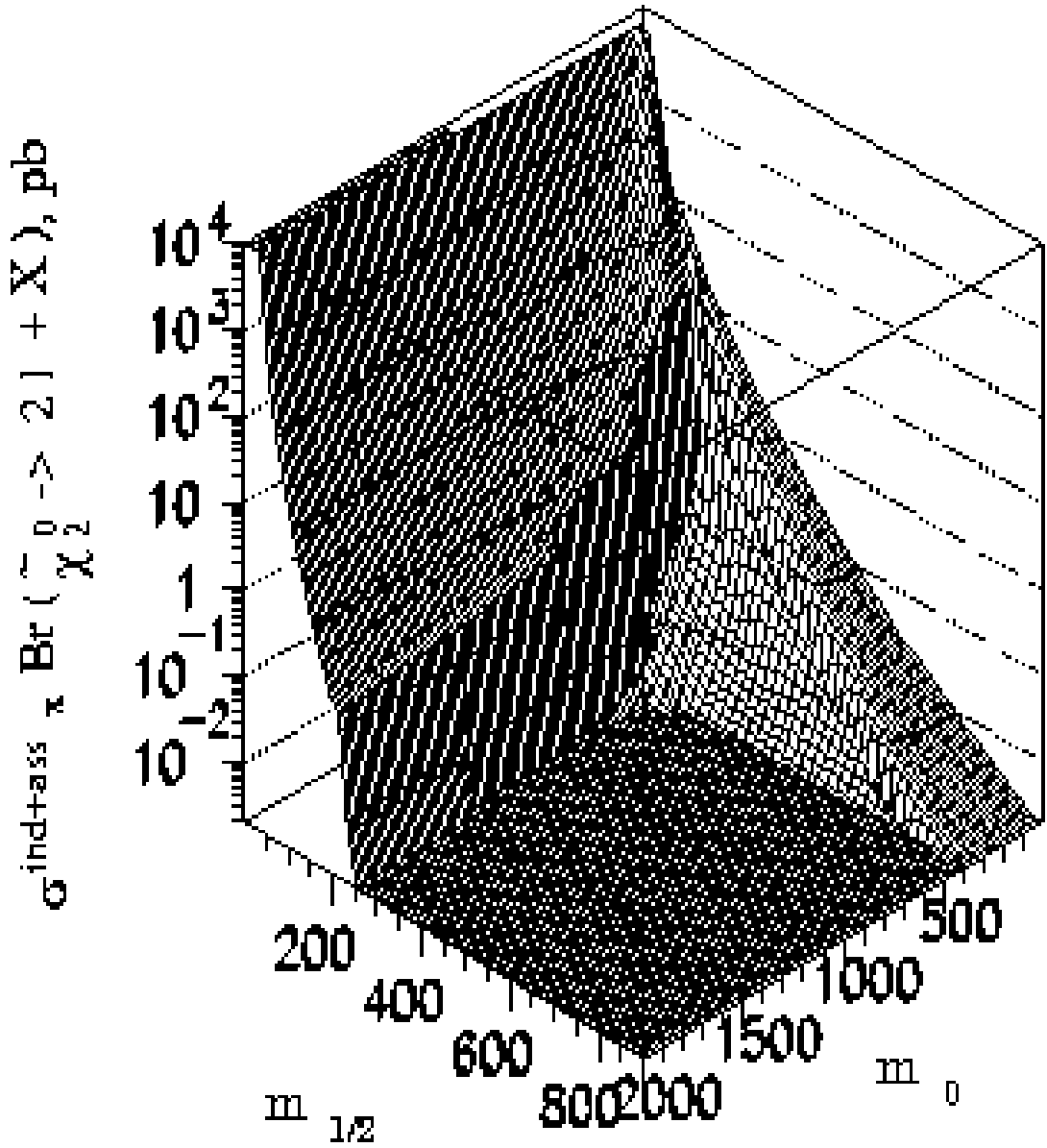}
\vspace{0.5cm}
\caption{Sigma times branching   
ratios for indirect and associated production of $\tilde{\chi}_2^0$ 
followed by decays into leptons as a function of $m_0$ and $m_{1/2}$.}
\end{center}
\end{figure}

\clearpage
\begin{figure}
\begin{center}
   \epsfig{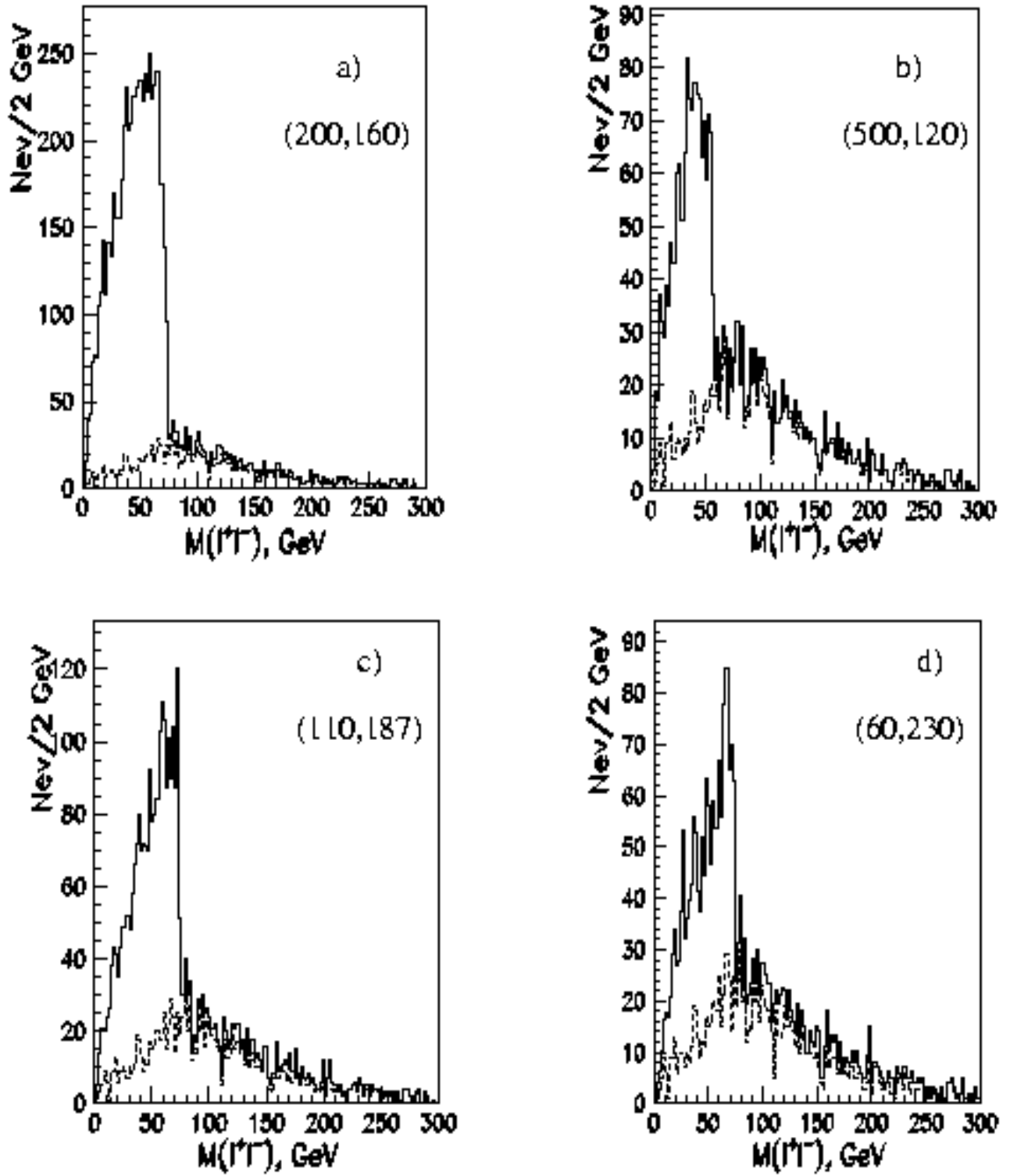}
\vspace{0.5cm}
\caption{Invariant mass distribution of two same--flavour, opposite--sign 
leptons at various 
$(m_0,m_{1/2})$ points from domain I, II and III for $L_{int}=10^3$ pb$^{-1}$.
SM background is also shown (dashed line).}
\end{center}
\end{figure}

\clearpage
\begin{figure}
\begin{center}
   \epsfig{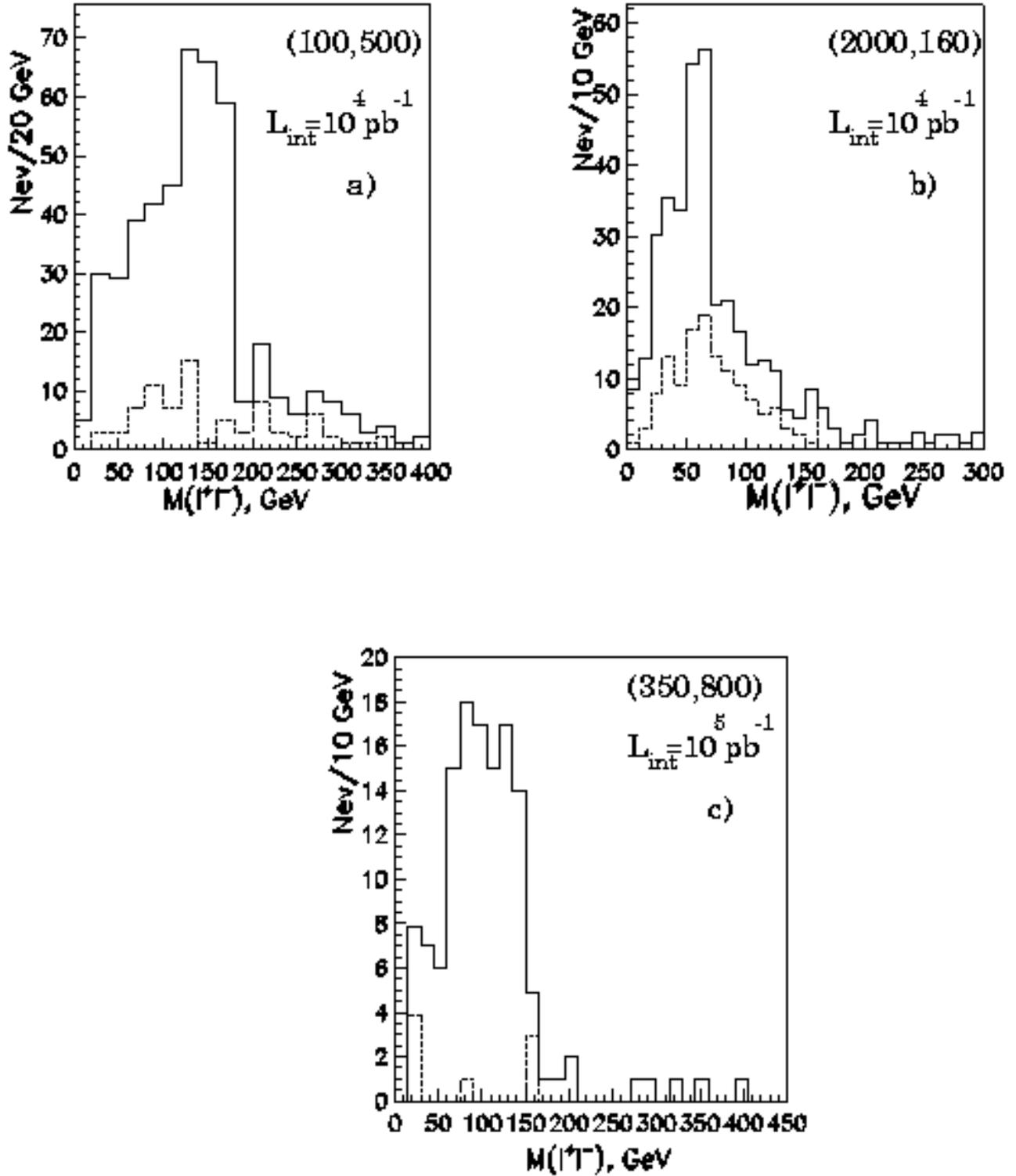}
\vspace{0.5cm}
\caption{Invariant mass distribution of two same--flavour, opposite--sign 
leptons at $(m_0,m_{1/2})$ 
points from domain I, II and III close to the experimental reach at 
corresponding luminosities $L_{int}=10^4$ pb$^{-1}$ and $10^5$ pb$^{-1}$.    
SM background is also shown  (dashed line).}
\end{center}
\end{figure}

\clearpage
\begin{figure}
\begin{center}
\vspace{-3cm}
   \epsfig{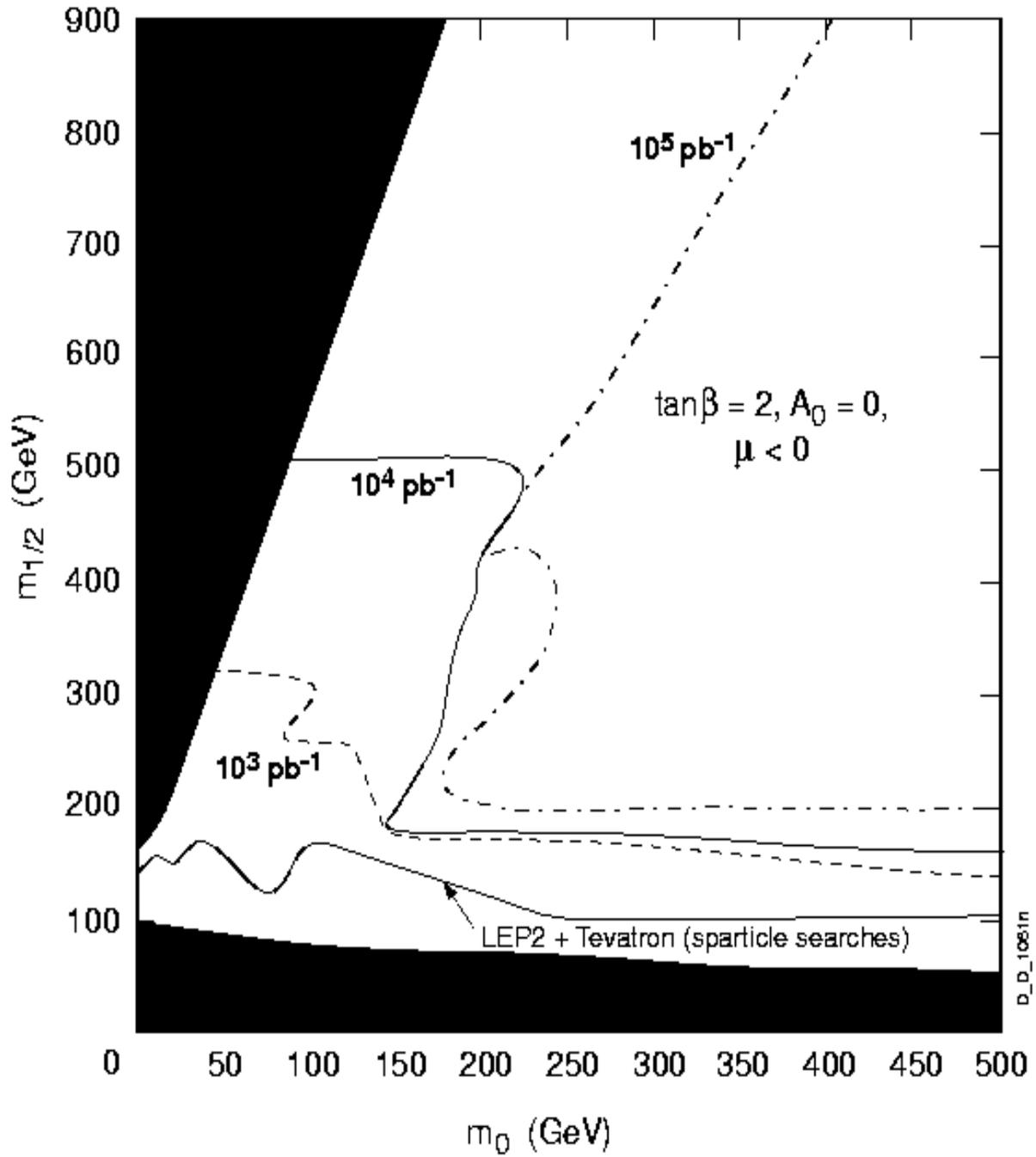}
\caption{Observability of edges in invariant dilepton mass 
distribution with luminosities $10^3$ pb$^{-1}$ (dashed line), 
$10^4$ pb$^{-1}$ (solid line) and $10^5$ pb$^{-1}$ (dashed-dotted line). 
Also shown are the explorable 
domain in sparticle searches at LEP2 (300 pb$^{-1}$) and the Tevatron 
(1 fb$^{-1}$), theoretically and experimentally excluded regions [19].}   
\end{center}
\end{figure}

\clearpage
\begin{figure}
\begin{center}
   \epsfig{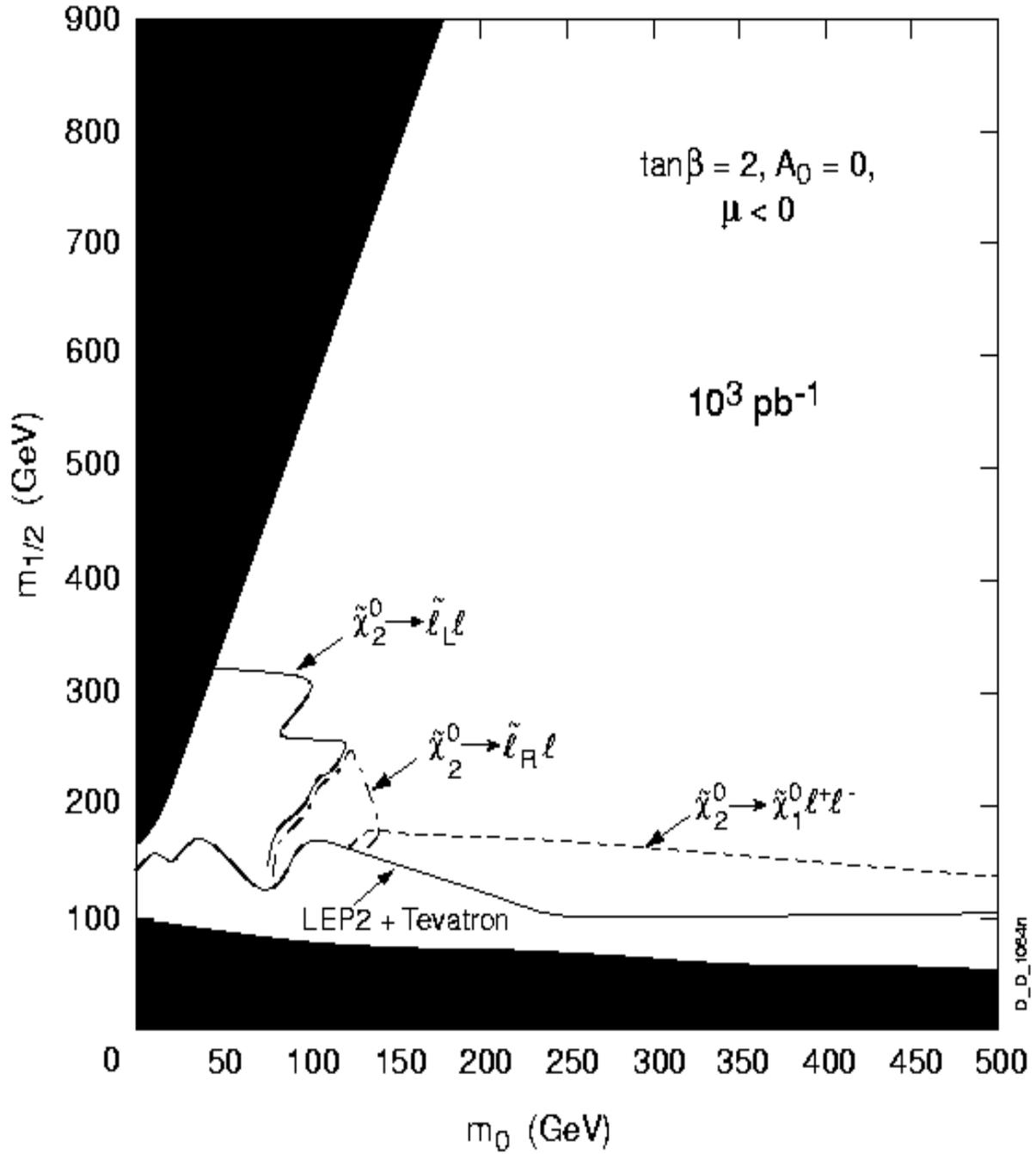}
\vspace{0.5cm}
\caption{ Domains where the observed  edge in the 
$M_{l^+l^-}$ distribution is due to the decays 
$\tilde{\chi}_2^0 \rightarrow \tilde{l}_L^{\pm} l^{\mp} \rightarrow 
\tilde{\chi}_1^0 l^+ l^-$  (solid line),  
$\tilde{\chi}_2^0 \rightarrow \tilde{l}_R^{\pm} l^{\mp} \rightarrow 
\tilde{\chi}_1^0 l^+ l^-$ (dashed-dotted line),
$\tilde{\chi}_2^0 \rightarrow \tilde{\chi}_1^0 l^+ l^-$ (dashed line),
$L_{int}=10^3$ pb$^{-1}$.}
\end{center}
\end{figure}

\clearpage
\begin{figure}
\begin{center}
   \epsfig{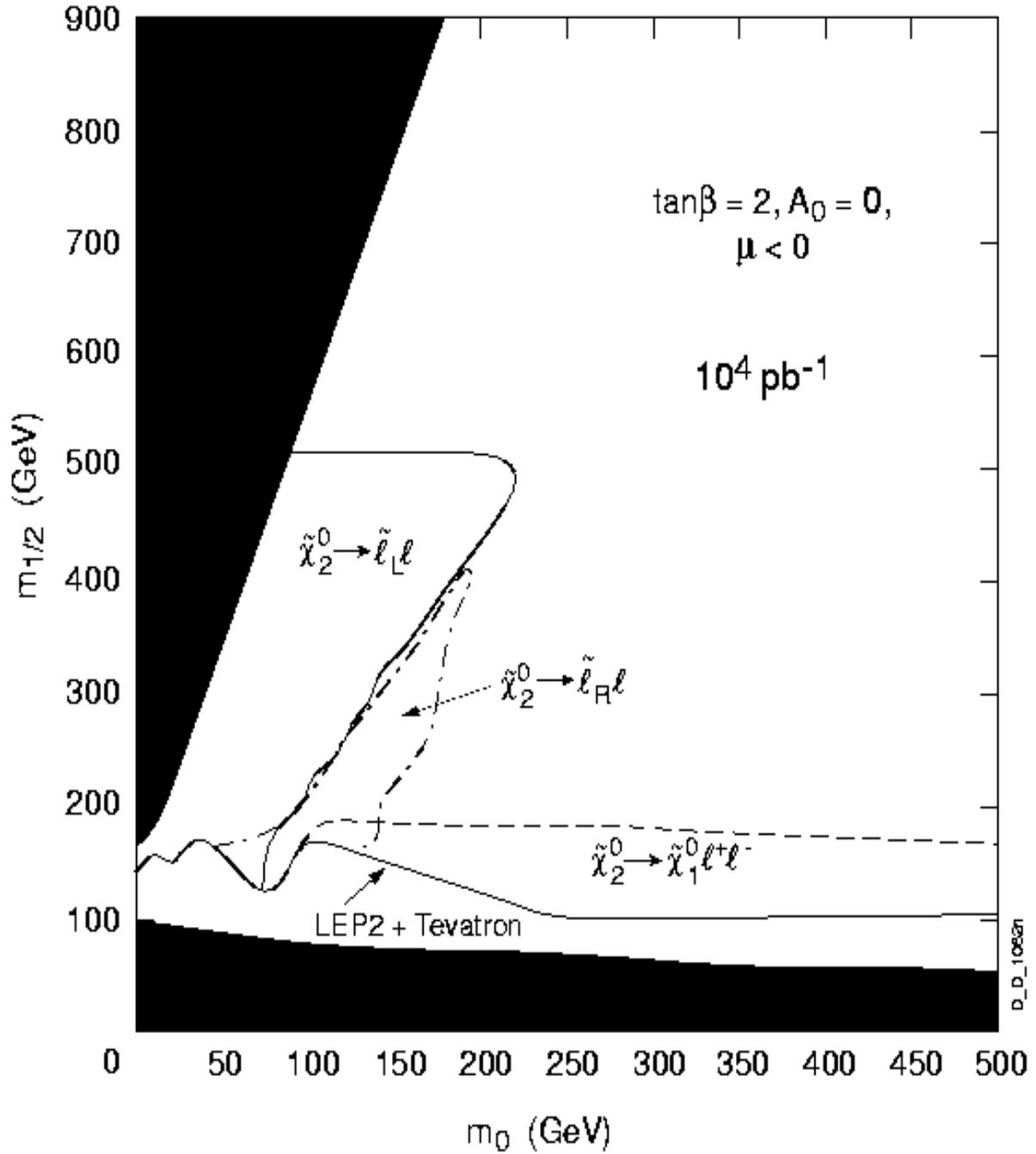}
\vspace{0.5cm}
\caption{
Domains where the observed edge in the 
$M_{l^+l^-}$ distribution is due to the decays 
$\tilde{\chi}_2^0 \rightarrow \tilde{l}_L^{\pm} l^{\mp} \rightarrow  
\tilde{\chi}_1^0 l^+ l^-$ (solid line),  
$\tilde{\chi}_2^0 \rightarrow \tilde{l}_R^{\pm} l^{\mp} \rightarrow 
\tilde{\chi}_1^0 l^+ l^-$ (dashed-dotted line),
$\tilde{\chi}_2^0 \rightarrow \tilde{\chi}_1^0 l^+ l^-$ (dashed line), 
$L_{int}=10^4$ pb$^{-1}$.}
\end{center}
\end{figure}

\clearpage
\begin{figure}
\begin{center}
   \epsfig{file=D_Denegri_1063n.pcx,width=17cm,height=18cm}
\vspace{0.5cm}
\caption{
Domains where the observed edge in the 
$M_{l^+l^-}$ distribution is due to the decays  
$\tilde{\chi}_2^0 \rightarrow \tilde{l}_L^{\pm} l^{\mp} \rightarrow 
\tilde{\chi}_1^0 l^+ l^-$ (solid line),  
$\tilde{\chi}_2^0 \rightarrow \tilde{l}_R^{\pm} l^{\mp} \rightarrow
\tilde{\chi}_1^0 l^+ l^-$ (dashed-dotted line),
$\tilde{\chi}_2^0 \rightarrow \tilde{\chi}_1^0 l^+ l^-$ (dashed line),
$L_{int}=10^5$ pb$^{-1}$.}
\end{center}
\end{figure}

\clearpage
\begin{figure}
\begin{center}
   \epsfig{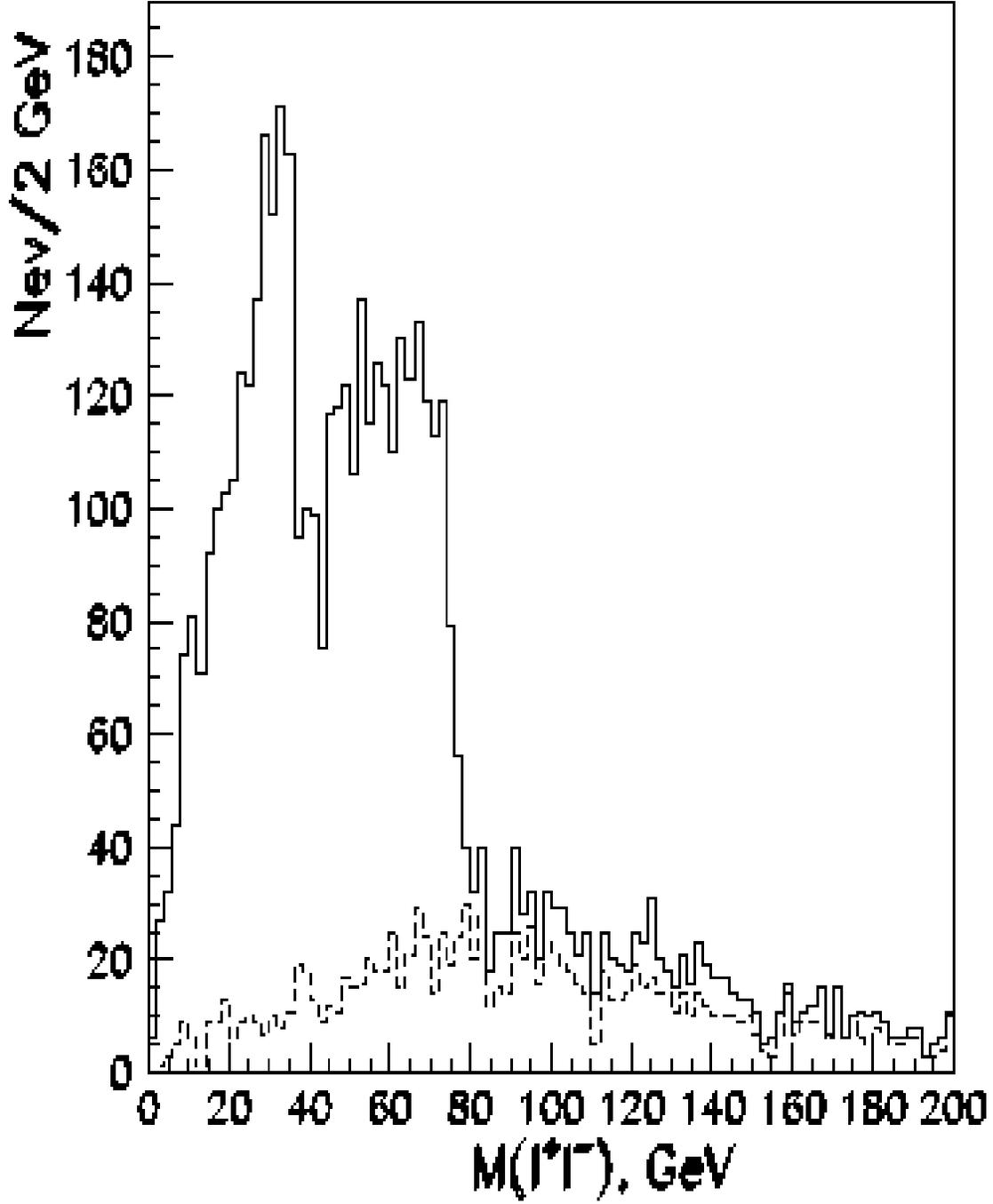}
\vspace{0.5cm}
\caption{Invariant mass distribution of two same--flavour, opposite--sign
leptons at the point $m_0=125$ GeV, $m_{1/2}=170$ GeV,
$L_{int}=10^3$ pb$^{-1}$. SM background is also shown (dashed line).
The edge at $M_{l^+l^-}=38$ GeV is due to the two--body
$\tilde{\chi}_2^0 \rightarrow \tilde{l}_R^{\pm} l^{\mp} \rightarrow
\tilde{\chi}_1^0 l^+ l^-$ decays and the edge at
$M_{l^+l^-}= 80$ GeV is due to  the three--body  
$\tilde{\chi}_2^0 \rightarrow  \tilde{\chi}_1^0 l^+ l^- $ decay.} 
\end{center}
\end{figure}

\clearpage
\begin{figure}
\begin{center}
   \epsfig{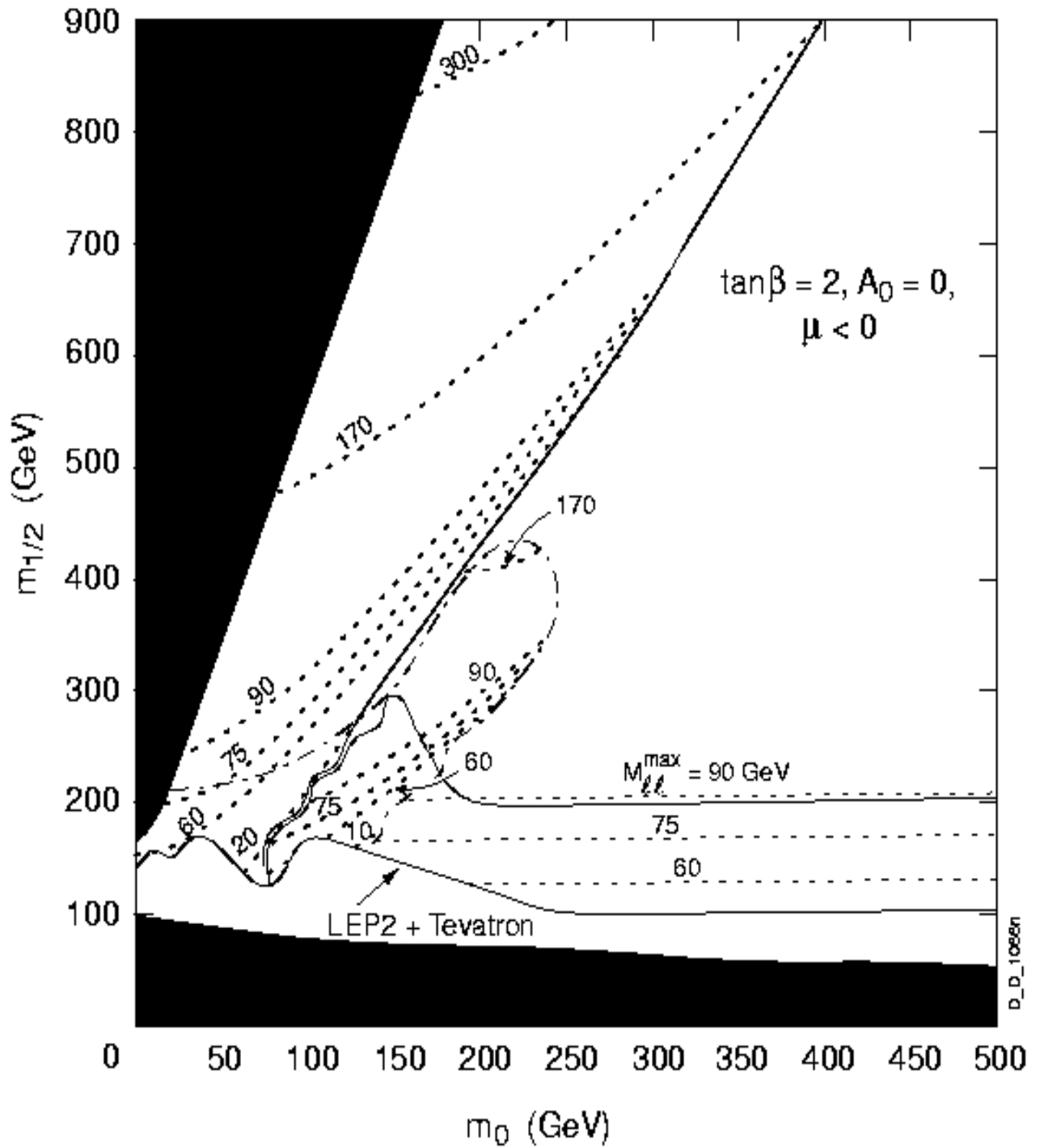}
\vspace{0.5cm}
\caption{Contour lines of expected $M_{l^+l^-}^{max}$ values (in GeV)
in the invariant dilepton mass distribution corresponding to the 
three different $\tilde{\chi}_2^0$ decay modes
in the region of the ($m_0,m_{1/2}$) parameter plane accessible with 
$10^5$ pb$^{-1}$.}
\end{center}
\end{figure}

\clearpage
\begin{figure}
\begin{center}
   \epsfig{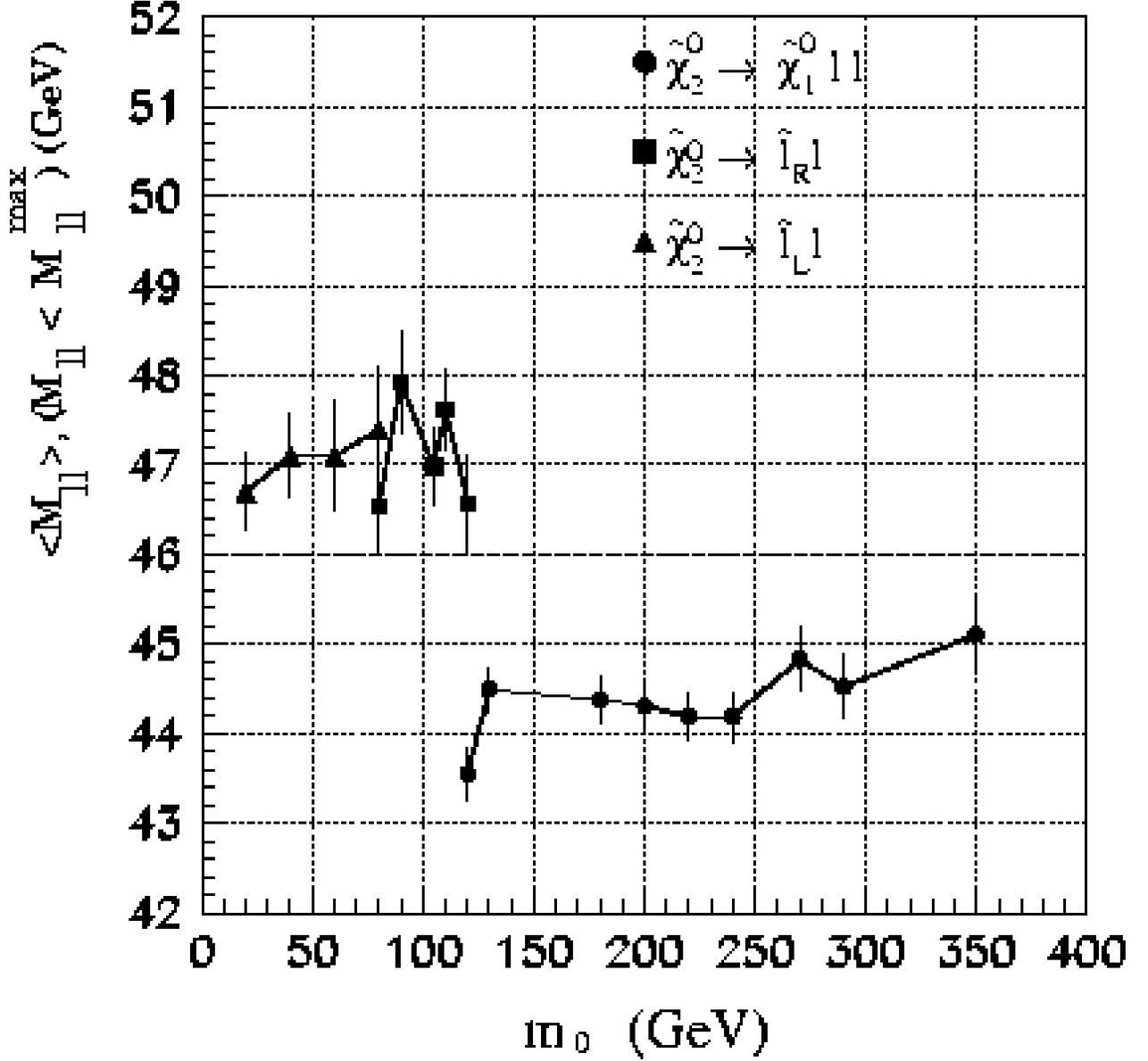}
\vspace{0.5cm}
\caption{Average value of invariant dilepton mass,
$M_{l^+l^-}<M_{l^+l^-}^{max}$ at ($m_0,m_{1/2}$) points with
$M_{l^+l^-}^{max}=74\pm 1$ GeV corresponding to the decays 
$\tilde{\chi}_2^0\rightarrow \tilde{\chi}_1^0 l^+ l^-$,
$\tilde{\chi}_2^0\rightarrow \tilde{l}_L^{\pm} l^{\mp} \rightarrow
\tilde{\chi}_1^0 l^+ l^-$ and
$\tilde{\chi}_2^0\rightarrow \tilde{l}_R^{\mp} l^{\mp} \rightarrow
\tilde{\chi}_1^0 l^+ l^-$  as a function of $m_0$; 
$p_T^{l_{1,2}}>15$ GeV, $E_T^{miss}>100$ GeV, $L_{int}=10^3$ pb$^{-1}$.}
\end{center}
\end{figure}

\clearpage
\begin{figure}
\begin{center}
   \epsfig{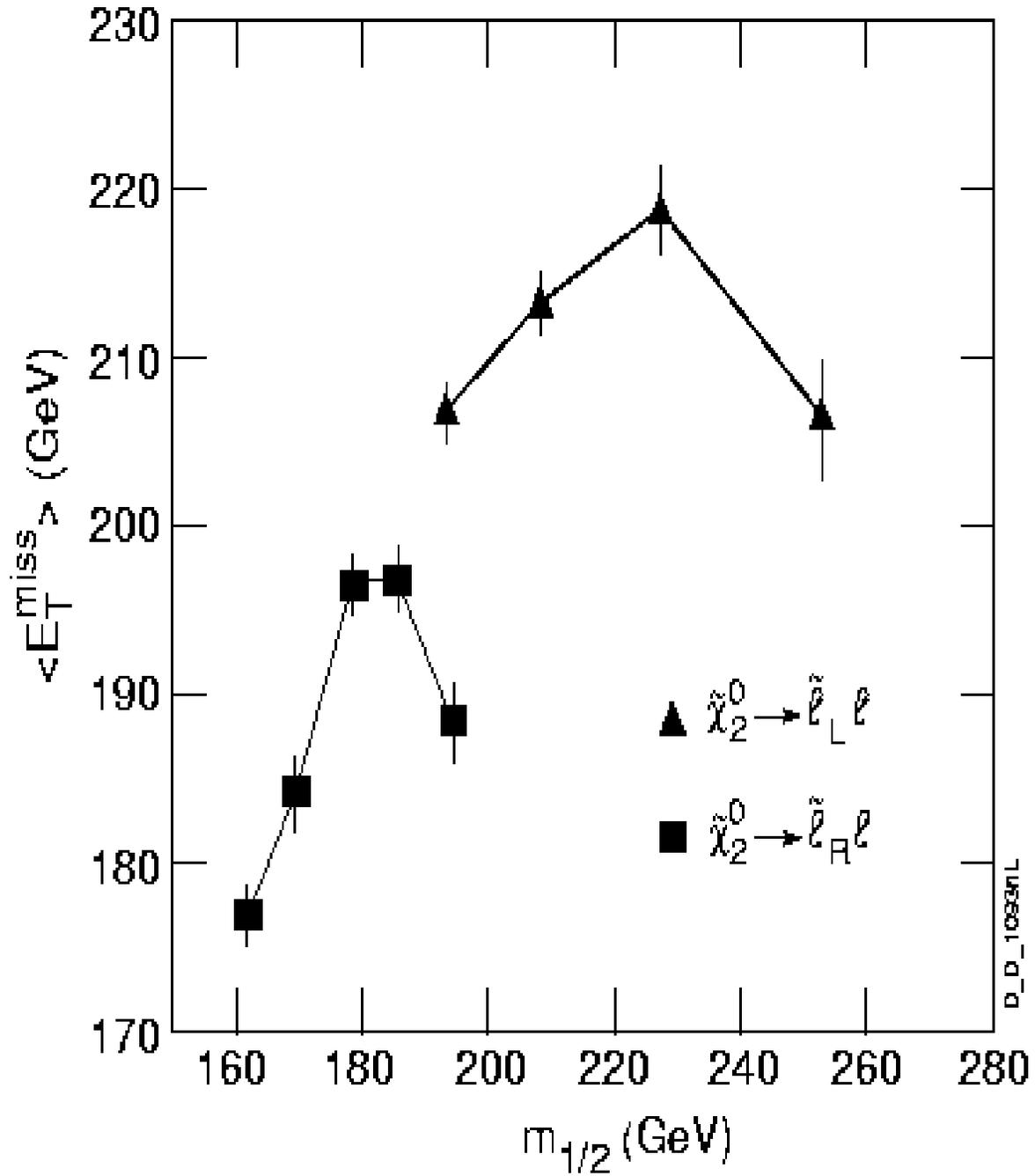}
\vspace{0.5cm}
\caption{Average value of $E_T^{miss}$ at ($m_0,m_{1/2}$) points with
$M_{l^+l^-}^{max}=74\pm 1$ GeV corresponding to the decays 
 $\tilde{\chi}_2^0\rightarrow \tilde{l}_L^{\pm} l^{\mp} \rightarrow
\tilde{\chi}_1^0 l^+ l^-$ and
 $\tilde{\chi}_2^0\rightarrow \tilde{l}_R^{\mp} l^{\mp} \rightarrow
\tilde{\chi}_1^0 l^+ l^-$ as a function of $m_{1/2}$, 
$L_{int}=10^3$ pb$^{-1}$. Event selection criteria are:
$p_T^{l_{1,2}}>15$ GeV, $E_T^{miss}>100$ GeV and 
$M_{l^+l^-}<M_{l^+l^-}^{max}$.}
\end{center}
\end{figure}

\clearpage
\begin{figure}
\begin{center}
   \epsfig{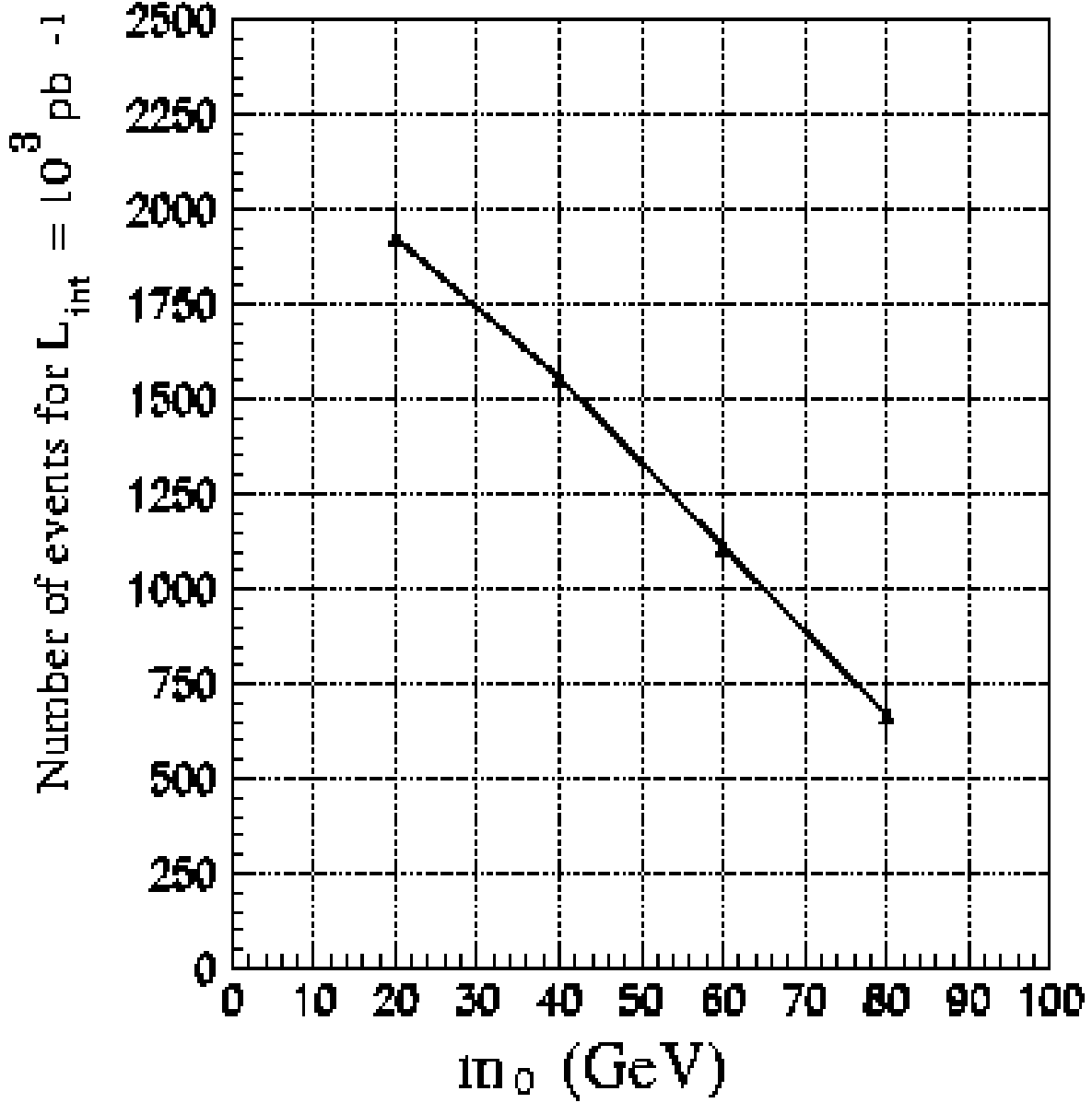}
\vspace{0.5cm}
\caption{Expected $2 l+E_T^{miss}$ event rate with $L_{int}=10^3$ pb$^{-1}$
along $M_{l^+l^-}^{max}=74\pm 1$ GeV contour line
in domain III ($\tilde{\chi}_2^0\rightarrow \tilde{l}_L^{\pm} l^{\mp}
\rightarrow \tilde{\chi}_1^0 l^+ l^-$) as a function of $m_0$. 
Event selection criteria are: 
$p_T^{l_{1,2}}>15$ GeV, $E_T^{miss}>130$ GeV and $M_{l^+l^-}<M_{l^+l^-}^{max}$
at corresponding points.} 
\end{center}
\end{figure}

\clearpage
\begin{figure}
\begin{center}
   \epsfig{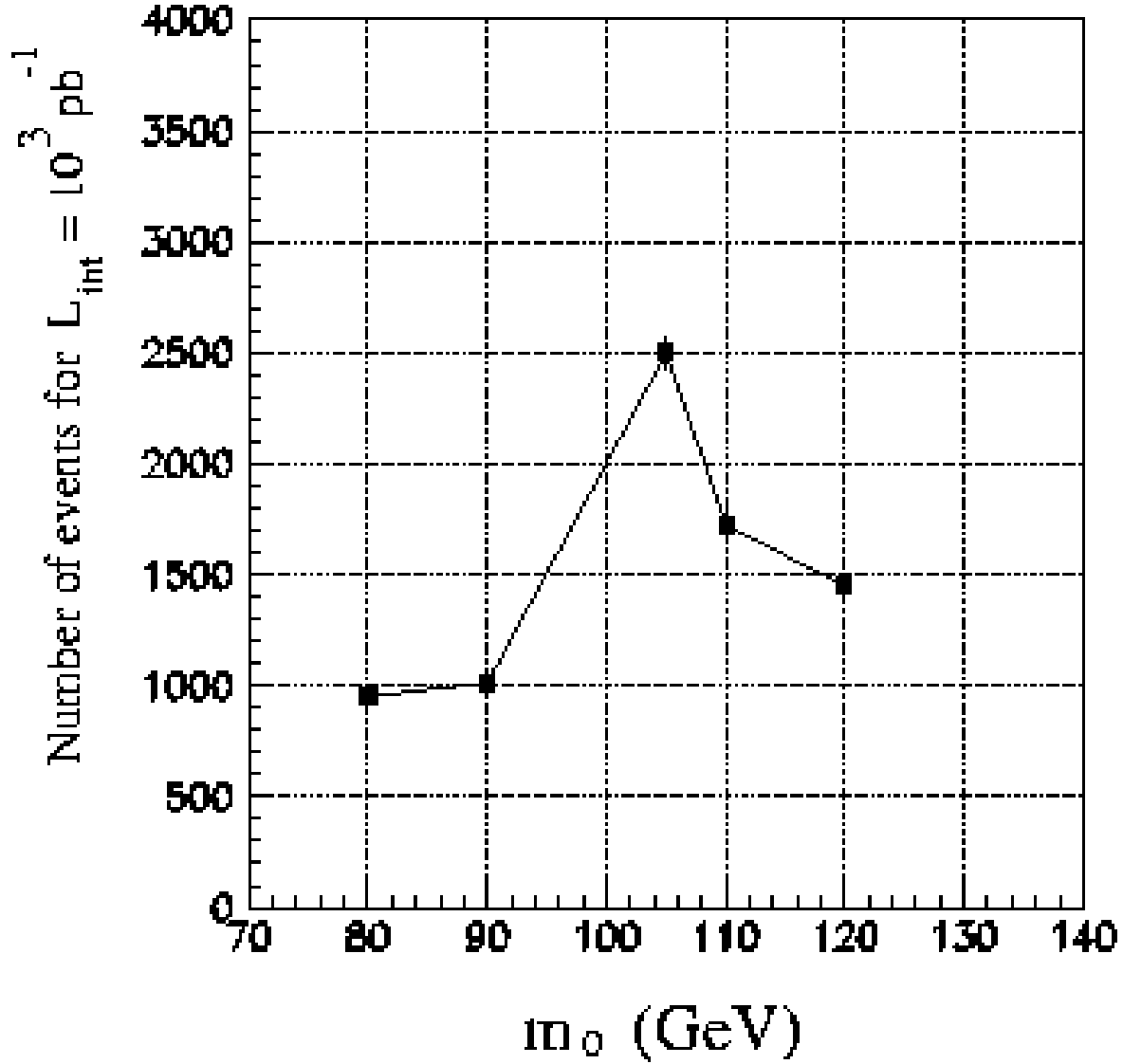}
\vspace{0.5cm}
\caption{Expected  $2 l+E_T^{miss}$ event rate with $L_{int}=10^3$ pb$^{-1}$
along $M_{l^+l^-}^{max}=74\pm 1$ GeV contour line in domain II 
($\tilde{\chi}_2^0\rightarrow \tilde{l}_R^{\pm} l^{\mp}
\rightarrow \tilde{\chi}_1^0 l^+ l^-$) as a function of $m_0$. 
Event selection criteria are: 
$p_T^{l_{1,2}}>15$ GeV, $E_T^{miss}>130$ GeV and $M_{l^+l^-}<M_{l^+l^-}^{max}$
at corresponding points.}
\end{center}
\end{figure}

\clearpage
\begin{figure}
\begin{center}
 \vspace{-1cm}  
\epsfig{file=stat-d1.pcx,width=13cm,height=22cm}
\caption{a) Expected  $2 l+E_T^{miss}$ event rate with $L_{int}=10^3$ pb$^{-1}$
along $M_{l^+l^-}^{max}=74\pm 1$ GeV contour line in domain I 
($\tilde{\chi}_2^0\rightarrow \tilde{\chi}_1^0 l^+ l^-$)
as a function of $m_0$.
Event selection criteria are: 
$p_T^{l_{1,2}}>15$ GeV, $E_T^{miss}>130$ GeV and $M_{l^+l^-}<M_{l^+l^-}^{max}$
at corresponding points; 
b) Average number of jets ($E_T^{jet}>30$ GeV, $\mid \eta_{jet} \mid <3$)
at investigated points from domain I. The numbers in parenthesis show the 
masses of squarks and gluinos (in GeV) at corresponding points.}
\end{center}
\end{figure}


\begin{thebibliography}{6}
\bibitem{ref1} Proc. of the ECFA Large Hadron Collider Workshop,
               Aachen, 1990, CERN 90-10, ECFA 90-133 (G.~Jarlskog, 
               P.~Rein,etc.);
             
               CMS Collaboration, Technical Proposal, LHCC/P2 (1994);

               ATLAS Collaboration, Technical Proposal, LHCC/P2 (1994).

\bibitem{ref2} H.~Baer, C-H.~Chen, F.~Paige and X.~Tata, 
  Phys. Rev. {\bf D 52}, 2746 (1995); Phys. Rev. {\bf D 53}, 6241 (1996);

  I.~Hinchliffe, J.~Womersley, LBNL-38997.

\bibitem{ref3} S.~Abdullin, CMS TN/96-095.
   
\bibitem{ref4} F.~del Aguila and Ll.~Ametller, Phys. Lett. {\bf B 261}, 
               326 (1991).


\bibitem{ref5} H.~Baer, C-H.~Chen, F.~Paige and X.~Tata, 
  Phys.Rev. {\bf D 49}, 3283 (1994).

\bibitem{ref6} D.~Denegri, L.~Rurua, N.~Stepanov, CMS TN/96-059.        

\bibitem{ref7} For reviews, see H.P.~Nilles, Phys. Rep. {\bf 110}, 1 (1984);

               P.~Nath, R.~Arnowitt and A.~Chamseddine, 
         Applied N=1 Supergravity, ICTP series in Theoretical Physics
         (World Scientific, Singapure, 1984); 

         M.~Drees and S.P.~Martin, hep-ph/9504324. 

\bibitem{ref8} L.~Rurua, presentations to CMS Collaboration Meetings, January
               and July 1996, and in the LHCC SUSY Workshop, CERN, October 
               29-30, 1996, CMS Document 1996-149 (PH-SUSY);
               
               D.~Denegri, L.~Rurua, N.~Stepanov, CMS TN/96-059. 

\bibitem{ref9} C.~Giunti, C.W.~Kim and U.W.~Lee, Mod. Phys. Lett.
               {\bf A 6}, (1991);

               J.~Ellis, S.~Kelley and D.V.~Nanopoulos, Phys.Lett. 
               {\bf B 260}, (1991) 161;

              U.~Amaldi, W.~de Boer and H.~Furstenau, Phys. Lett. 
              {\bf B 260}, 447 (1991);

              P.~Langacker and M.~Luo, Phys. Rev. {\bf D 44}, 817 (1991).


\bibitem{ISAJET} F.~Paige and S.~Protopopescu, in {\em Supercollider Physics},
  p. 41, ed. D.~Soper (World Scientific, 1986); 

  H.~Baer, F.~Paige, 
  S.~Protopopescu and X.~Tata, in {\em Proceedings of the Workshop on Physics
  at Current Accelerators and Superolliders}, ed. J.~Hewett, A.~White and
  D.~Zeppenfeld (Argonne National Laboratory, 1993).

\bibitem{edges} H.~Baer, K.~Hagiwara, X.~Tata, Phys. Rev. {\bf D 35}, 1598 
                (1987);
             
    H.~Baer, D.D.~Karatas, X.~Tata, Phys. Rev. {\bf D 42}, 2259 (1990) 
        (fig. 6a);

    H.~Baer, C.~Kao, X.~Tata, Phys. Rev. {\bf D 48}, 5175 (1993); 

    H.~Baer, C.-H.~Chen, F.~Paige, X.~Tata, Phys. Rev. {\bf D 50}, 4508 (1994).

\bibitem{Snowmass} 
F.~Paige, Determining SUSY particle masses at LHC, Proc. of 
the 1996 DPF/DPB Summer Study on High-Energy Physics
"New Directions for High-Energy Physics", Snowmass, Colorado, 1996, p.710;
 
A.Bartl et al., Supersymmetry at LHC, ibidem, p.693;

J.~Amundson et al., Report of the Supersymmetry Theory
Subgroup, ibidem, p.655.

\bibitem{13} CMS presentation at the LHCC SUSY Workshop, CERN, October 29-30,
               1996, CMS Document 1996-149 (PH-SUSY).
               
                  
\bibitem{sleptons} K.~Inoue, A.~Kakuto, H.~Komatsu, and S.~Takeshita, 
                     Prog. Theor. Phys. {\bf 68}, 927 (1982).

\bibitem{CMSJET} S.~Abdullin, A.~Khanov and N.~Stepanov, CMS TN/94-180.


\bibitem{muons} V.~Karimaki, CMS TN/94-151.


\bibitem{hadrons} V.~Genchev, L.~Litov, Study of the CMS TP calorimetry system,
                  CMS TN/94-272 (1994).

\bibitem{PYTHIA} T.~Sj\"{o}strand,
  {\em Comp. Phys. Com.} {\bf 39}, 347 (1986);

  T.~Sj\"{o}strand and M.~Bengtsson,
  {\em Comp. Phys. Com.} {\bf 43}, 367 (1987);

  H.U.~Bengtsson and T.~Sj\"{o}strand,
  {\em Comp. Phys. Com.} {\bf 46}, 43 (1987);

  T.~Sj\"{o}strand, CERN-TH.7112/93.

\bibitem{ref19} H.~Baer, M.~Brhlik, Phys. Rev. {\bf D 53}, 597 (1996). 
\end{thebibliography}
\end{document}